\documentclass[onecolumn]{aastex701}

\usepackage{comment}

\pdfoutput=1

\received{4 December, 2025}
\revised{2 March, 2026}
\accepted{-- --, 2025}

\shorttitle{ICME evolution from 0.07--5.4 AU}
\shortauthors{C. Möstl et al.}
\graphicspath{{./}{figures/}}

\begin{document}

\title{On the magnetic field evolution of interplanetary coronal mass ejections from 0.07 to 5.4 au}

\author[0000-0001-6868-4152]{Christian Möstl}
\affiliation{Austrian Space Weather Office, GeoSphere Austria, Graz, Austria}
\email{christian.moestl@geosphere.at}
\correspondingauthor{C. Möstl}

\author[0000-0001-9992-8471]{Emma E. Davies}
\affiliation{Austrian Space Weather Office, GeoSphere Austria, Graz, Austria}
\email{emma.davies@geosphere.at}

\author[0009-0004-8761-3789]{Eva Weiler}
\affiliation{Austrian Space Weather Office, GeoSphere Austria, Graz, Austria}
\affiliation{Institute of Physics, University of Graz, Graz, Austria}
\email{eva.weiler@geosphere.at}

\author[0000-0002-2559-2669]{Hannah T. Rüdisser}
\affiliation{Austrian Space Weather Office, GeoSphere Austria, Graz, Austria}
\affiliation{Institute of Physics, University of Graz, Graz, Austria}
\email{hannah.ruedisser@geosphere.at}

\author[0000-0003-1516-5441]{Ute V. Amerstorfer}
\affiliation{Austrian Space Weather Office, GeoSphere Austria, Graz, Austria}
\email{ute.amerstorfer@geosphere.at}

\author[0000-0002-6273-4320]{Andreas J. Weiss}
\affiliation{Goddard Planetary Heliophysics Institute, University of Maryland, Baltimore County, Baltimore, MD 21250, USA}
\affiliation{Heliophysics Science Division, NASA Goddard Space Flight Center, Greenbelt, MD, USA}
\email{ajefweiss@gmail.com}

\author[0000-0002-6362-5054]{Martin A. Reiss}
\affiliation{Community Coordinated Modeling Center, NASA Goddard Space Flight Center, Greenbelt, MD 20771, USA}
\email{martin.reiss@nasa.gov}

\author[0000-0002-6553-3807]{Satabdwa Majumdar}
\affiliation{Austrian Space Weather Office, GeoSphere Austria, Graz, Austria}
\email{satabdwa.majumdar@geosphere.at}

\author[0000-0002-7572-4690]{Timothy S. Horbury}
\affiliation{Imperial College London, South Kensington Campus, London SW7 2AZ, UK}
\email{t.horbury@imperial.ac.uk}

\author[0000-0002-1989-3596]{Stuart D. Bale}
\affiliation{Physics Department and Space Sciences Laboratory, University of California, Berkeley, USA}
\email{bale@berkeley.edu}

\author[0000-0001-7894-8246]{Daniel Heyner}
\affiliation{Technical University of Braunschweig, Braunschweig, Germany}
\email{d.heyner@tu-braunschweig.de}




\begin{abstract}
A central question for understanding interplanetary coronal mass ejection (ICME) physics and improving  space weather forecasting is how ICMEs evolve in interplanetary space. We have updated one of the most comprehensive in situ ICME catalogs to date, which now includes 1976 events from 11 space missions covering over 34 years, from December 1990 to August 2025. We have combined existing catalogs including magnetic obstacles and identified and added boundaries of an additional 807 (40.8\%) events ourselves. With this catalog, we demonstrate the most extensive analysis to date of total ICME magnetic field values as a function of heliocentric distance. Parker Solar Probe has observed 6 ICMEs at $< 0.23$~au (until April 2025), and Solar Orbiter and BepiColombo have added more events near 0.3~au, bridging the major observational gap towards the solar corona. Our main result is that a single power law can describe the evolution of the mean total magnetic field (exponent value of $k=-1.57$) and maximum field ($k=-1.53$) for ICMEs with magnetic obstacles (MOs), from 0.07 to 5.4~au. Extending the power law to the solar photosphere reveals a strong inconsistency with magnetic field magnitudes observed in the quiet Sun and active regions by 2 and 4 orders of magnitude, respectively. We introduce a multipole-type power law with two exponents, $k_1=-1.57$, and $k_2=-6$, relating the ICME magnetic field magnitude to an average solar active region field strength. These results present important observational constraints for the evolution of ICMEs from the Sun to the heliosphere.
\end{abstract}



\section{Introduction} \label{sec:intro}


Decades of observations of interplanetary coronal mass ejections (ICMEs) have still left us with many unsolved puzzles. Some of the key ones among them are the 3D magnetic structure of the flux rope and its extent in all three dimensions. Furthermore, how they evolve after their launch from the Sun towards interplanetary space, passing the planets and moving towards the outer heliosphere is still not fully understood \citep[e.g.][]{manchester2017, luhmann_2020_review,temmer2023, alhaddad2025structure}. With the Parker Solar Probe \cite[PSP;][]{Fox_2016} mission, launched in 2018, a new region below 0.3~au has been unlocked for the first time for in situ observations of ICMEs \cite[e.g.][]{nieves_chinchilla_2020,winslow_2021_sta_psp,salman_2024_psp_cat, davies2024, trotta2024, jebaraj2024}. 

Here, we combine these in situ observations of ICMEs by PSP with data from ten other spacecraft missions to form a catalog that is as uniform as possible. Our ICMECAT catalog contains 1976 ICMEs observed in situ, covers 3 solar cycles from 1990 to 2025, and heliocentric distances from 0.07--5.4 au. This means we present the most comprehensive analysis to date of the radial evolution of the total magnetic field strength of ICMEs measured in situ. We put a particular emphasis on their magnetic obstacle intervals, which often contain magnetic flux ropes \citep[e.g.][]{cane2003} that drive strong geomagnetic storms with their potential long-duration southward magnetic fields \citep[e.g.][]{zhang2007}. This information is of major importance for developing and validating empirical and numerical models of ICME flux ropes for many heliophysics applications. 

Many authors have considered power law fits for the relationship between the total magnetic field in ICMEs against distance from the Sun \citep[e.g.,][]{leitner_2007, gulisano2007, good2016, Hanneson2020, davies2021_juno, davies2022, salman_2024_psp_cat}. Statistical results indicate power laws with exponents $k$ in the range between $k=-1$ and $k=-2$, depending on the heliocentric distance range covered. However, studies that use multi-spacecraft conjunctions for the same event at different heliocentric distances paint a picture where individual events can differ from the average parameters of the power laws \citep[e.g.][]{leitner_2007,Salman_2020,davies2022}. For example, \cite{Davies_2021_solo} demonstrate a case of almost no radial expansion between 0.8 and 1.0~au, as the slow CME that they study is pushed by a high-speed stream from behind and thus constrained its expansion. Even with individual events deviating from these empirical relationships, the laws are a highly useful basis to describe the physics of CME evolution with empirical models.

Our intent with this study is not only to improve the understanding of the physics of CMEs, but the results are highly useful for practical space weather applications, as there is now renewed interest in the concept of so-called upstream or sub-L1 monitors \citep[see][]{lindsay1999,cyr2000diamond,kubicka2016,laker2024, lugaz2025need, weiler2025, davies2026_solo}. These are spacecraft that measure the magnetic field strength at a position closer to the Sun than the Sun--Earth Lagrange 1 (L1) point, but still near the Sun--Earth line, providing earlier measurements of the ICME magnetic field structure than current space weather monitors at L1, extending the forecast lead time. The European Space Agency (ESA) is currently building the HENON spacecraft \citep[e.g.][]{cicalo2025henon}, planned to be launched at the end of 2026. This mission acts as a pathfinder for ESA’s proposed sub-L1 mission concept, SHIELD, which plans to deploy multiple spacecraft on a distant retrograde orbit \citep{henon1969,perozzi2017} at heliocentric distances of about 0.85–0.9 au. Power-law scaling of ICME magnetic fields provides a straightforward method to extrapolate the observations of these sub-L1 spacecraft to Earth distance.

The total magnetic field magnitude as well as the north-south $B_z$ component of a CME arriving at Earth is a major factor in its geo-effectiveness. Therefore, many empirical models must account for the evolution of the CME magnetic field as it propagates from the Sun to the Earth \citep[e.g.][]{isavnin_2016, moestl_2018,weiss_2021_fit,pal2022_model,sarkar2024, weiss2024distorted}. The low heliocentric distances of the recent PSP observations now allow us to connect the resulting power laws for the total magnetic field to the magnetic field values observed in the quiet Sun photosphere and active regions. The rationale behind this is that semi-empirical CME flux rope models should ideally be driven by the magnetic field strengths measured in active regions and then reproduce the correct magnetic field values at larger heliocentric distances as the CME propagates away from the Sun \citep[e.g.][]{pomoell2019active}. In the case of the 3DCORE flux rope model \citep{moestl_2018, weiss_2021_solo, ruedisser_2024, davies2024}, the evolution of the total magnetic field is prescribed by a power law, the exponent of which currently has to be constrained using in situ observations, typically from spacecraft at the Sun--Earth L1 point, or potentially from future sub-L1 monitors. Ultimately, our aim is to determine a power-law relation that allows us to start from the magnetic field strength of the active region at the photosphere (1$R_{\odot}$) and model the magnetic field of the CME throughout the heliosphere.

Power-law descriptions of ICME magnetic fields are also crucial for Farady rotation methods, which use radio observations to remotely probe the magnetic structure of ICMEs \citep[e.g.][]{wood2020faraday, kooi2022_faraday,jensen2025polarized}. These methods require assumptions about how the magnetic field decreases with heliocentric distance. Therefore, accurate power laws are essential for converting observed Faraday rotation into predictions of magnetic field strengths at larger heliocentric distances. This is particularly important as Faraday rotation offers one of the few options to obtain information about a CMEs internal magnetic field structure before it is measured in situ.

Our ICMECAT catalog \citep{Moestl2017,moestl2020} has grown over the recent years into a unified and comprehensive database of in situ ICMEs observed by 11 spacecraft over more than 3 solar cycles. It has been widely used in different contexts, not only for case studies, but also for studying intermittency in CMEs \citep{ruohotie2025}, for training machine learning algorithms for automatic identification of ICMEs in solar wind in situ observations \citep{nguyen2019,ruedisser2022, ruedisser2026arcane}, or for removing ICMEs to study the radial evolution of solar wind parameters \citep{yogesh2026}. Its main purpose is to identify multi-spacecraft observations of CMEs to advance our knowledge of their 3D evolution, shape and structure \citep[e.g.][]{moestl2022,regnault2024,davies2024,palmerio2025event4, alhaddad2025structure,zhang2025_multipoint}.

\section{Spacecraft observations} \label{sec:data}

\begin{figure*}
\centering
{\includegraphics[width=\textwidth]{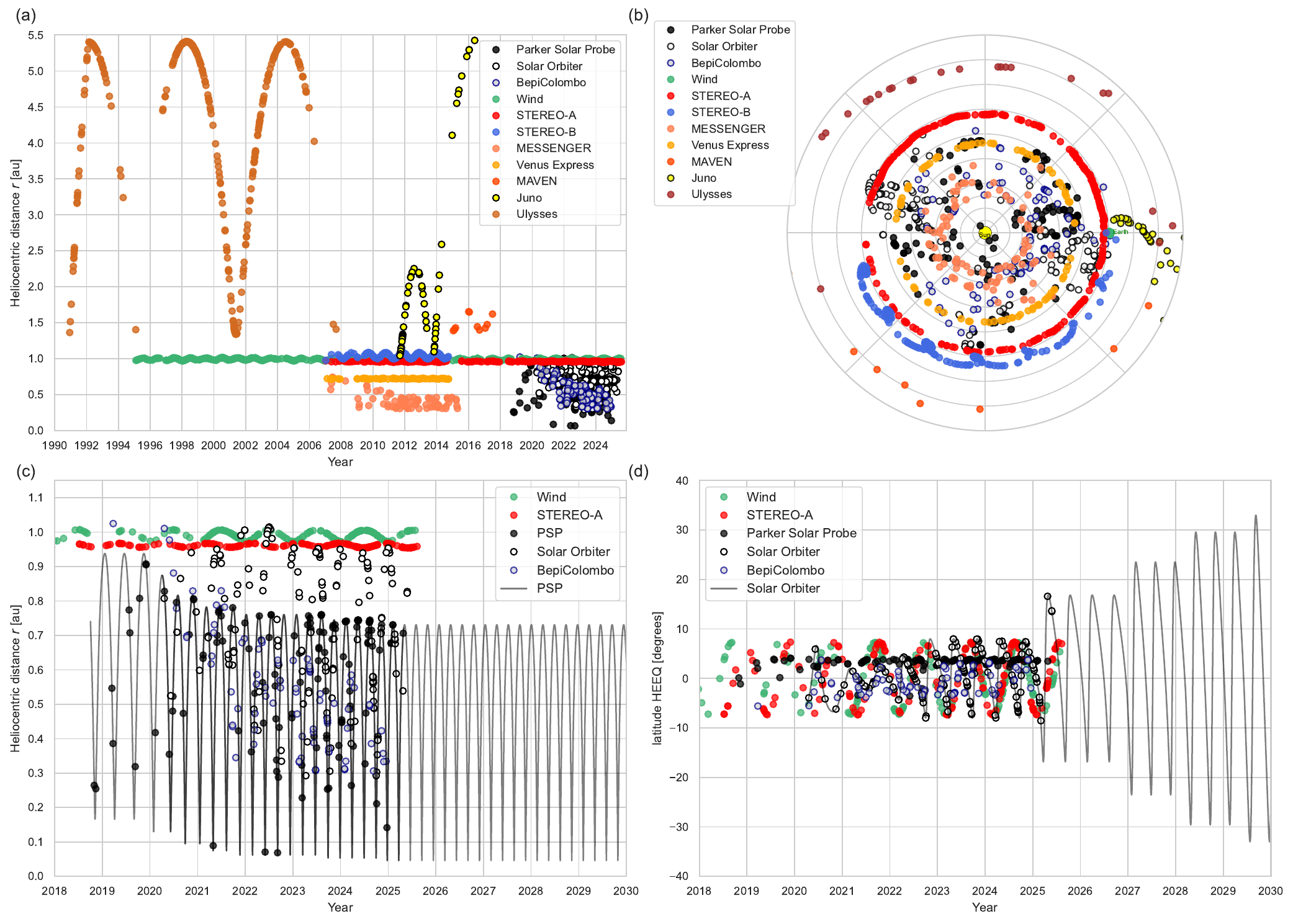}}
\caption{Overview of ICME observations. (a) Heliocentric distance at which all events in the ICMECAT catalog were observed over time. Each dot represents an ICME observation, with colors corresponding to the different spacecraft by which they were observed in situ. (b) Longitude and radial distance (HEEQ) of all spacecraft positions at time of ICME observation, up to 1.6~au. (c) Heliocentric distance coverage of ICME events observed in the inner heliosphere ($\sim$~$<$~1~au) since the launch of PSP in 2018, with an extension of the PSP orbit up to 2030 (gray line). (d) Coverage in latitude (HEEQ) of ICME events observed in the inner heliosphere from 2018 onwards with the Solar Orbiter trajectory (gray line) extended up to 2030.} 
\label{fig:1_stats}
\end{figure*}

\begin{figure*}
\centering
{\includegraphics[width=\textwidth]{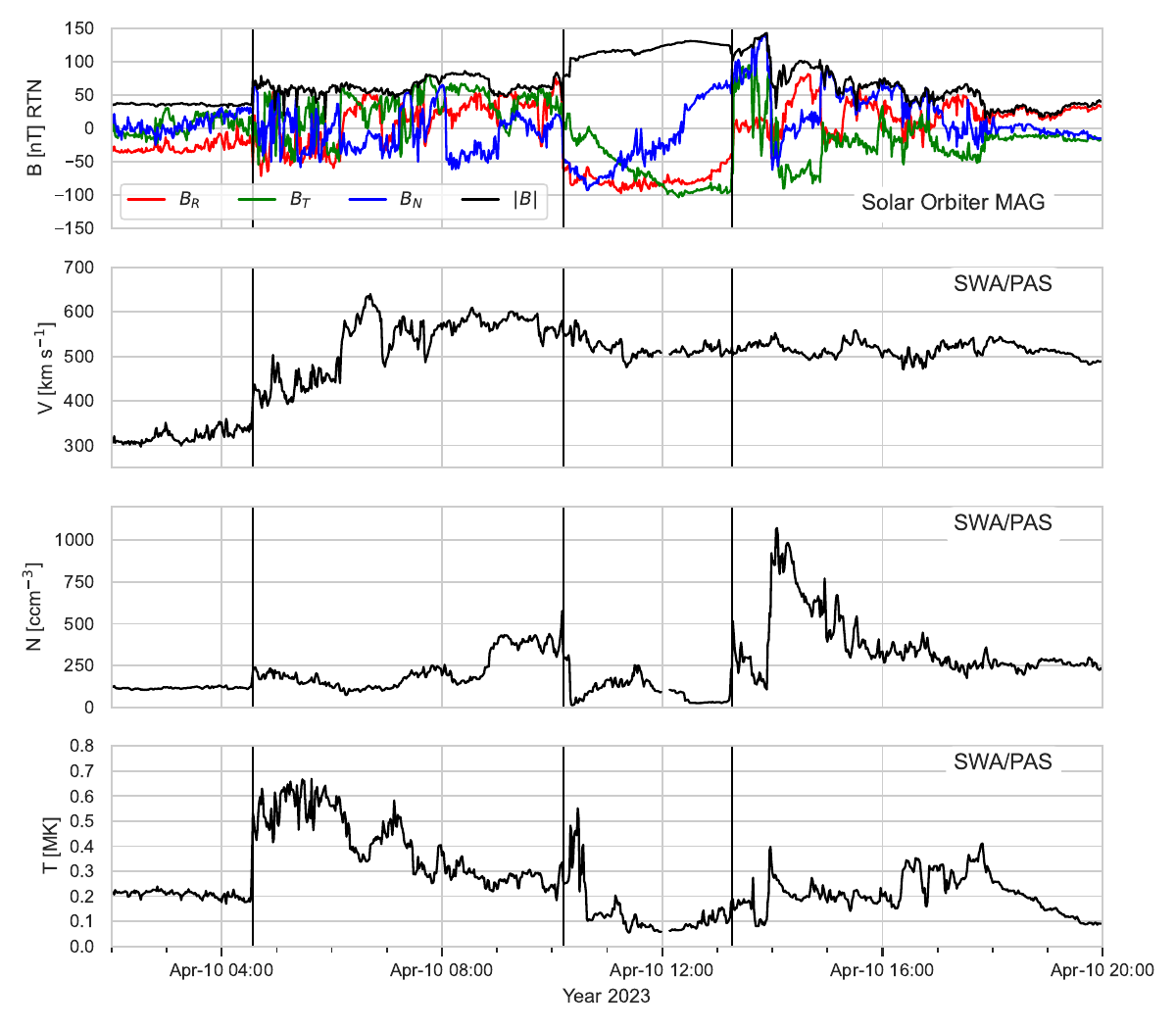}}
\caption{An example in situ observation of an ICME in the presented ICMECAT catalog: Solar Orbiter observed this event at a minimum distance of 0.29~au on 2023 April 10. From left to right, the vertical lines state the shock arrival time and the beginning and end times of the magnetic obstacle, which in this case is a well defined magnetic flux rope of south-east-north type. The panels from top to bottom show magnetic field components (in RTN coordinates) and total field, proton bulk speed, proton density, and proton temperature. } 
\label{fig:2_solo}
\end{figure*}


\begin{figure*}
\centering
{\includegraphics[width=\textwidth]{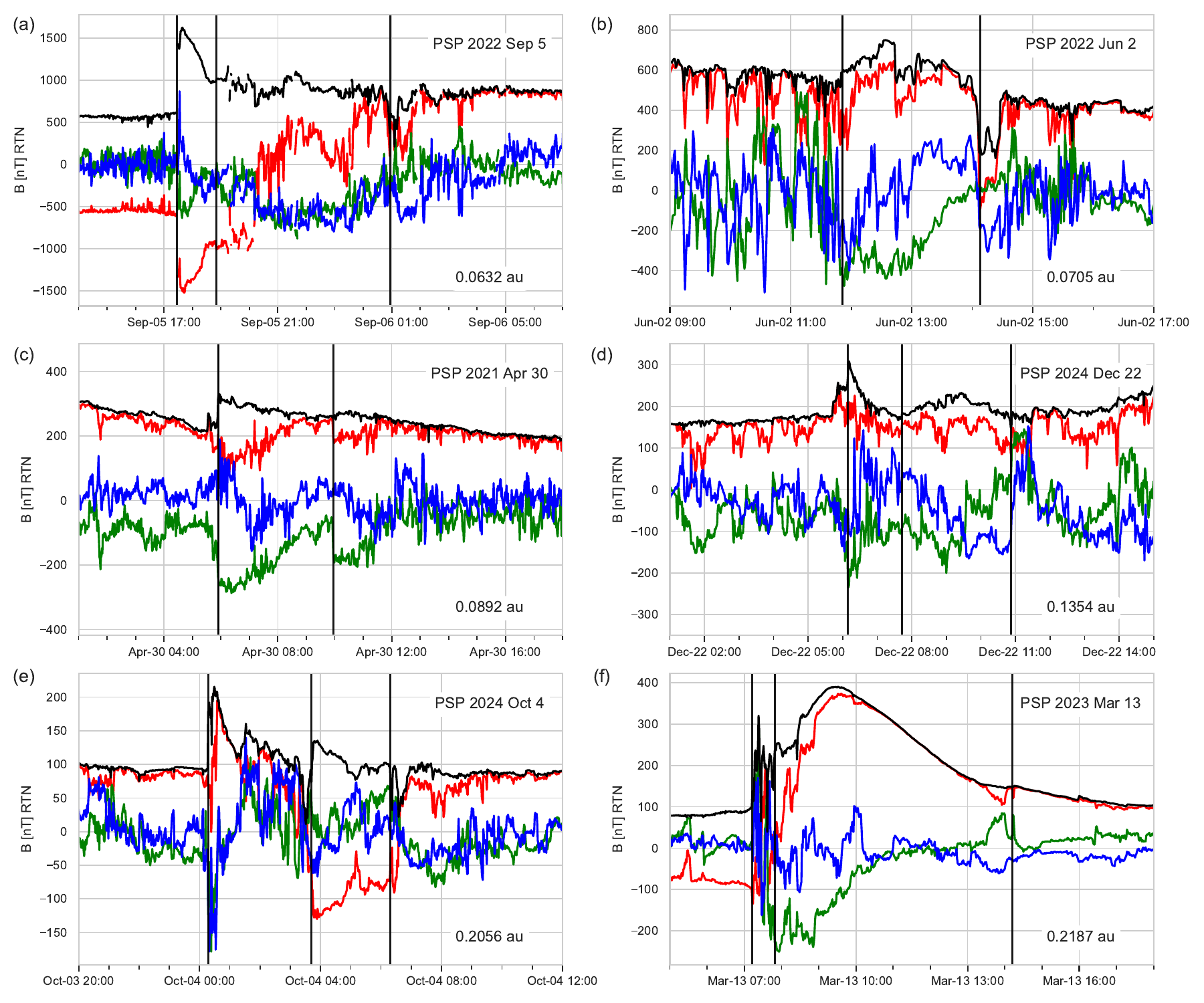}}
\caption{In situ CME observations close to the Sun: The total magnetic field and components of all six ICMEs observed by PSP/FIELDS until 2025 April 30 at distances $< 0.23$~au, starting with the closest event in 2022 September 5. Each panel shows the magnetic field in RTN coordinates, with the event start date and the minimum heliocentric distance PSP reached during the full ICME observation (between ICME start time and MO end time). From left to right, the vertical black lines delineate the ICME start time, the MO start time and MO end time. When there are only two vertical lines in panels (b) and (c), they mark the MO start time and MO end time.} 
\label{fig:3_psp}
\end{figure*}

\subsection{ICME catalog}

At the heart of our analysis is the ICMECAT, an open source and living ICME catalog which is accessible online (see Data Availability Statement), presented so far in \cite{Moestl2017} and \cite{moestl2020}. A search for multipoint ICME events using the ICMECAT has been demonstrated in \cite{moestl2022}. Here, we present a major update that now acts as the main reference publication for the ICMECAT, and briefly summarize its main features. We have significantly extended the catalog in comparison to previous work, with the main goal of adding more ICME events observed in the inner heliosphere ($< 1$~au) that have recently become available, possible due to the coverage of more recent spacecraft missions. Our aim is to provide a living database to better understand CME magnetic field evolution, which rapidly includes novel events soon after they are observed. With respect to the solar activity in solar cycle 25, rising from solar minimum around the launches of PSP in 2018 and Solar Orbiter in 2020 to solar maximum around 2024, the updated catalog now includes ICMEs for about half of solar cycle 25. It also covers the first 23 orbits of PSP until 2025 April 30, of which two orbits reached the planned closest approach distance of the mission (0.0458~au).

Figure~\ref{fig:1_stats}a shows the heliocentric distance at which the ICMECAT events were observed over time, where the different color markers correspond to the observing spacecraft. The ICMECAT now covers 1976 ICME events from the years 1990 to 2025. Figure~\ref{fig:1_stats}b gives an overview of the longitudinal distribution of events in Heliocentric Earth Equatorial (HEEQ) coordinates against radial distance, where the Earth is fixed on the right side at 0° longitude. This demonstrates how many ICMEs have been observed in the space age after 1990, in particular around 1~au by the STEREO mission and through several missions venturing into the inner heliosphere, including both planetary spacecraft (e.g. MESSENGER, Venus Express, and BepiColombo) and dedicated solar wind missions (e.g. PSP and Solar Orbiter). Figures~\ref{fig:1_stats}c and \ref{fig:1_stats}d show a subset of events with the (predicted) trajectories of PSP and Solar Orbiter, respectively, as well as the more recent ICME events, with PSP observing events at distances $<0.3$~au at almost every recent close approach to the Sun. BepiColombo is measuring ICMEs with magnetic field instrument MPO-MAG \citep{heyner2021} until orbit insertion at Mercury in 2026, and Solar Orbiter is now starting to provide high latitude ICME observations since raising its orbital inclination after February 2025.



\begin{table}[ht]
\centering
\caption{Source catalogs included in the ICMECAT, with time range, number of events, spacecraft. ICMECAT as reference means these are our own identifications.}
\label{tab:catalog_sources}
\begin{tabular}{lllll}
\hline
Catalog / Reference & Time Range & \# Events & Spacecraft \\
\hline

ICMECAT & 1998 April -- 2025 August & 209 & Wind  \\ 
\cite{nieves_chinchilla_2020} &  1995 February -- 2015 December &  342 & Wind  \\

ICMECAT & 2017 July -- 2025 August & 175 & STEREO-A  \\ 

\cite{jian2018} & 2007 January -- 2017 December & 188 & STEREO-A  \\

\cite{jian2018} & 2007 January -- 2014 September & 150 & STEREO-B  \\

\cite{davies2021_juno, davies2022} & 2011 September -- 2016 May & 53 & Juno  \\

ICMECAT & 2012 June -- 2014 September & 13 & VEX  \\
\cite{good2018} & 2007 February -- 2013 December & 80 & VEX  \\

ICMECAT & 2009 August -- 2015 April & 10 & MESSENGER  \\
\cite{good2018} & 2007 May -- 2011 November  & 22 & MESSENGER \\
\cite{winslow2018} & 2011 May -- 2014 September & 55 & MESSENGER \\

ICMECAT & 2015 January - 2017 July & 10 & MAVEN \\

\cite{richardson2014identification} & 1990 December -- 2007 September & 279 & Ulysses \\

ICMECAT &  2019 March -- 2024 December & 86 & BepiColombo \\

ICMECAT & 2018 October -- 2025 April & 145 & PSP \\

ICMECAT &  2020 April -- 2025 May  & 159   & Solar Orbiter \\

\hline
\end{tabular}\label{tab1:data}
\end{table}



The ICMECAT is an aggregate of catalogs published by \cite{nieves_chinchilla_2020} for Wind, \cite{jian2018} for STEREO-A/B, \cite{davies2021_juno, davies2022} for Juno, \cite{good2018} for Venus Express and MESSENGER, \cite{winslow2018} for MESSENGER and \cite{richardson2014identification} for Ulysses. In addition, we have added ICME events identified by ourselves for the missions STEREO-A, Wind, PSP, Solar Orbiter,  BepiColombo and a few events for MAVEN, hence including altogether 11 different spacecraft missions in the ICMECAT. We note that we have also adjusted some event times originally taken from the existing STEREO-A and Wind mission catalogs. In total, 807 events (40.8\% of the catalog) were included in addition to existing catalog events or modified by our team. In Table~\ref{tab1:data} we summarize the catalog sources.

When combining different catalogs, one needs to be careful with the criteria that were used to identify ICME events. While the catalogs that we use do not match perfectly in their criteria, all catalogs that we have used include magnetic obstacles (MOs, also known as magnetic ejecta). Shock arrivals, followed by sheath regions but lacking MOs, are not included in the ICMECAT, however, events with only MOs and no sheaths are included. All events were selected visually according to the criteria given below, namely for both the events given in the existing catalogs and those identified by ourselves.
Automatic identification methods \citep[e.g.][]{camporeale2017, nguyen2019, ruedisser2022, nguyen2025_multiclass, ruedisser2026arcane} are now reaching a promising level, but for consistency, we stick to manual identification for the moment.

All MOs in the ICMEs are broadly characterized by a high total magnetic field strength and a smooth rotation of the magnetic field vector. If plasma data is present, we look for additional signatures such as declining speed profiles and possibly low proton temperature within the MO, which are signatures associated with expansion of the MO. Each event has three times assigned to it: (1) ICME start time, (2) MO start time, and (3) MO end time. For interacting ICMEs, we try to identify individual MOs as standalone events \citep[e.g. for the May 2024 ICMEs, see][]{weiler2025}. A summary of the identification criteria is given in Table~\ref{tab:criteria}.

\begin{table}[h!]
\centering
\caption{Summarized criteria for interplanetary coronal mass ejection (ICME) identification.}
\label{tab:criteria}
\begin{tabular}{lll}
\hline
\textbf{Boundary} & \textbf{Observable Signature} & \textbf{Certainty} \\
\hline\hline
ICME start time & Shock (sudden jumps in magnetic field, speed, density, temperature) & high (shocks)  \\
                & or start of density enhancement; if none, similar to MO start time & average (density only) \\
\hline
MO start time   & Magnetic field data: start of elevated field strength, smooth rotation of & low, transition of sheath to  \\
             & field vectors over several hours, low field variance; look for discontinuities &flux rope often unclear \\
             & Plasma data: Start of low proton temperature, low proton density,  &\\
             & linear declining speed profile &  \\
\hline                
MO end time     & Magnetic field data: end of elevated field strength, smooth rotation of & average, certainty of  \\ 
                & field vectors over several hours, low field variance; look for discontinuities &  boundary for transition to  \\
                & Plasma data: End of low proton temperature, low proton density, & ambient solar wind varies\\
                & linear declining speed profile &  \\
\hline
\end{tabular}
\end{table}

If no shock is present, in our catalog the ICME start time reflects the beginning of a significant density enhancement in front of the MO. If neither is present, the ICME start time is set equal to the MO start time. It is well known that different human observers often disagree on the exact times of the MO interval and the ICME start time (if no shock is present). Nevertheless, these uncertainties that are inherently present in these types of catalogs will have only little influence on the derivation of the global statistical results that we present here, although we note that distributions of general parameters at 1~au can be affected \citep[see][]{kay2026collection}.



Concerning the in situ data, we use the magnetic field components and total field, as well as bulk proton data (speed, temperature, density, if available), linearly interpolated to a 1 minute resolution, in order to calculate parameters and plot a figure for each event in a computationally efficient way. For missions where the available cadence is greater than 1 minute, we use a different time resolution. For Wind, we interpolate the data to 2 minutes. For Ulysses, an 1~hour combined plasma and magnetic field dataset is readily available.  For the few MAVEN events that we include, we interpolate to one data point per orbit. This results in a largely homogeneous dataset for all 11 spacecraft with a total size of only about 10 gigabyte. The catalog is then created by reading in all available spacecraft data, and another file containing the heliospheric positions of each spacecraft produced using the SPICE kernels available from most missions (see figshare links in the data section at the end of this study). We use the spacecraft position files to provide a figure with an overview of the distribution of the various spacecraft in the inner heliosphere at the time each of the ICME events is observed (see the ICMECAT figure database on the website and on figshare). We then calculate up to 37 parameters for each ICME based only on the 3 boundary times defined for each event (ICME start, MO start and MO end). In this way, all parameters are fully consistently recalculated every time the catalog is updated. We do not rely on any parameters from any previously published catalog, except for the 3 times that define an event, and calculate all cataloged ICME parameters by ourselves.

\subsection{An example ICME observed by Solar Orbiter}

Figure~\ref{fig:2_solo} presents an example of an ICME identified in our catalog using Solar Orbiter magnetic field \citep[MAG instrument,][]{horbury_2020} and proton data \citep[Solar Wind Plasma Analyser (SWA) instrument,][]{owen2020}. Here, the ICME start time is given by the shock arrival time, defined by the first data point of an increase in proton speed, density, temperature, and total magnetic field. The MO start time marks the beginning of the MO interval, which in this case is clearly visible in the Solar Orbiter example as the magnetic field starts to rotate smoothly and the total field jumps to higher values. The proton density and temperature are lower during the MO interval for this event, which is often the case due to expansion. However, in this example, the expected speed signatures are not so obvious, and the density and temperature changes do not perfectly match the magnetic field changes, in particular at the beginning of the MO. 

This event serves to demonstrate that our approach is to focus on the magnetic obstacle as identified in the magnetic field data, because magnetometer data are available for all spacecraft, and the magnetic field data often give clear indications of the boundary times by the discontinuities that are present. The fact that the derived plasma times are not always coincident with the magnetic field signatures add to the subjectivity in setting ICME boundaries. From our experience, for many events, the front boundary of the MO is often not so clear, and different catalogs can differ in setting this time \citep{kay2026collection}. Sometimes the MO end time is more confidently identifiable than the MO start time, before a transition to the ambient wind occurs.

\subsection{Parker Solar Probe events}

A major novel data source we explore in this study are events observed close to the Sun by the NASA PSP spacecraft, of which we consider magnetic field \citep[FIELDS;][]{bale2016_fields} and bulk solar wind plasma \citep[Solar Wind Electrons Alphas and Protons, SWEAP;][]{kasper2016_sweap} data. Up to 2025 April 30, which is the end of the PSP/FIELDS dataset we have available at the time of writing, 12 ICMEs have been observed at distances $< 0.29$~au, and 6 at $< 0.23$~au. Such close distances to the Sun have never before been reached by any other mission.

Figure~\ref{fig:3_psp} shows the magnetic field observations measured by the FIELDS instrument for these 6 events. The closest CME event observed in situ so far is observed on 2022 September 5 (Figure~\ref{fig:3_psp}a), subject of many studies \citep[e.g.][]{romeo2023_sep2022,paouris2023_sep2022,davies2024,trotta2024, riley2025_sep2022}. The CME observed on 2022 June 2 (Figure~\ref{fig:3_psp}b) was studied by \cite{braga2024_june2022}, and the shock of an event on 2023 March 13 was analyzed in \cite{jebaraj2024}. In addition to these known events already discussed in the literature, we have added another three novel events, which we briefly discuss below.

Figure~\ref{fig:3_psp}c shows the magnetic field data from an event on 2021 April 30, which was observed at 0.0892~au and consists of a shock, with a jump in speed from $<200$~km~s$^{-1}$ to $>300$~km~s$^{-1}$ (not shown), and a magnetic obstacle directly following the shock. It is clearly connected to an eastward directed CME appearing at 2021 April 30 00~UT in SOHO/LASCO/C3. This CME impact happened right after the solar flyby in which PSP first entered below the critical Alfvén surface on 2021 April 28 \citep{kasper2021alfven}.


The ICME on 2024 December 22 (Figure~\ref{fig:3_psp}d) is a weaker event with respect to the general rise in total magnetic field strength, which nonetheless shows a sharp discontinuity in the total field followed by a longer elevation throughout the magnetic obstacle, during which the bipolar $B_N$ component changes sign from north to south and the $B_T$ component is unipolar, consistent with a low inclination flux rope \citep{bothmer1998}. 

The ICME observed by PSP on 2024 October 4 (Figure~\ref{fig:3_psp}e) consists of a shock followed by an MO, at 0.2056~au. The components are somewhat irregular in the MO, with a stronger $B_R$ component, signaling a non direct-hit. We are very confident that this is an ICME as it is also observed by Solar Orbiter at 0.30~au, only 4 hours later and 15° degrees longitude away. 


Finally, the event on 2023 March 13 (Figure~\ref{fig:3_psp}f) is one of the clearest ICME examples of this subset, along with the 2022 September 5 and 2022 June 2 events. PSP reached a minimum heliocentric distance of 0.2187~au during the ICME observations. It features a very clear shock and strongly elevated total magnetic field in the MO, although it includes a high $B_R > 0 $ component, signaling a more flank-like encounter of PSP with the ICME structure \citep[the shock was studied by][]{jebaraj2024}.

These near-Sun events are scientifically valuable not only because they represent the closest in situ ICME signatures to date, but also because they enable detailed case studies under conditions that are vastly different from those at 1~au. 
We now proceed using the full ICMECAT database to derive new power laws for the main CME magnetic field parameters, closing the gap between the solar corona and 0.3~au using the ICMEs observed by PSP, and a much larger set of events at further heliocentric distances.

\section{Results} \label{sec:results}


\begin{figure*}
\centering
{\includegraphics[width=\textwidth]{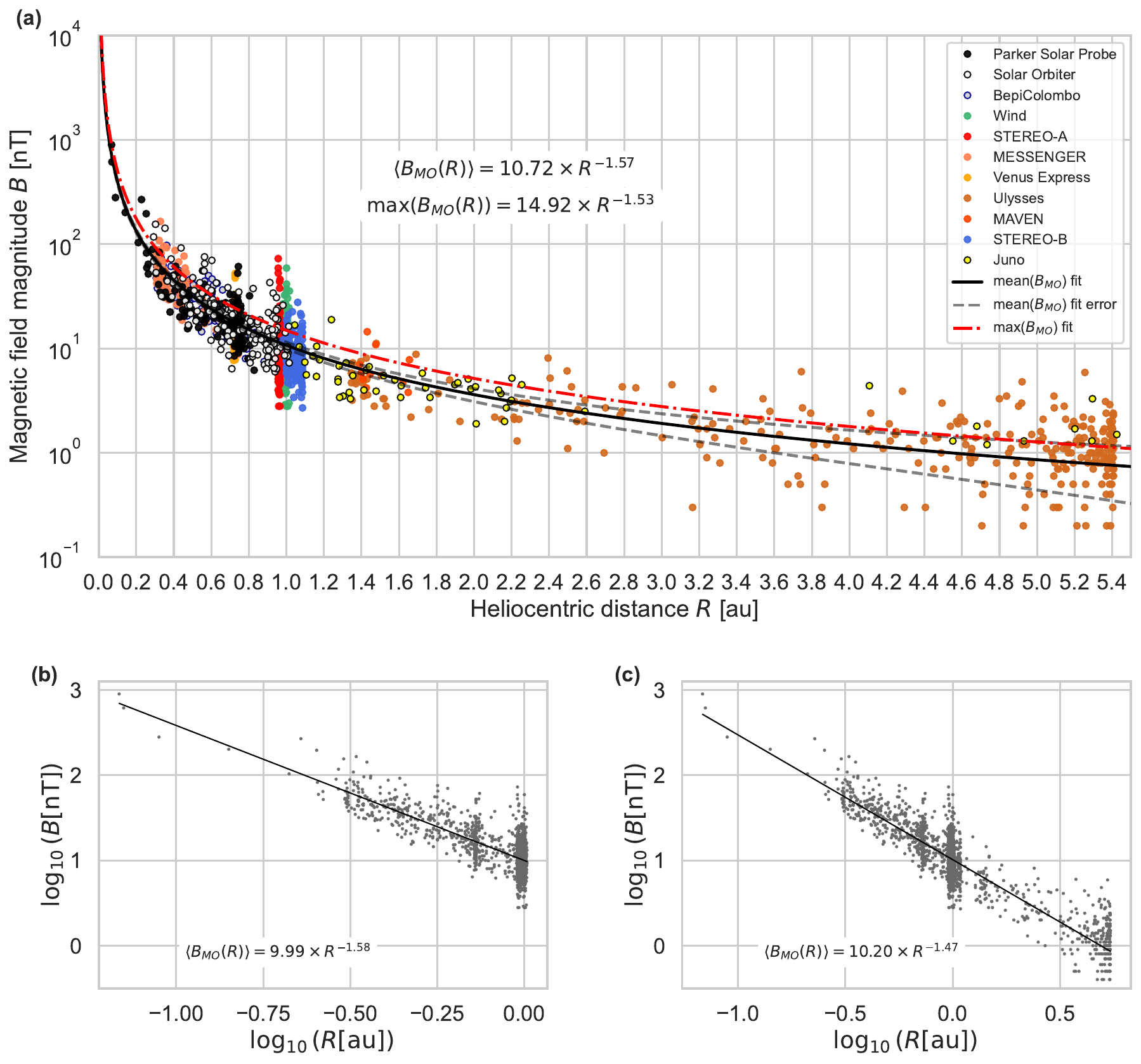}}
\caption{Evolution of ICME total magnetic field strength in their magnetic obstacles with increasing heliocentric distance. (a) Each dot represents the mean magnetic field of the magnetic obstacle $\langle B_{MO}(R) \rangle$ of an observed ICME by one of the 11 spacecraft indicated in the legend. The solid black line is a direct power law fit to $\langle B_{MO}(R) \rangle$ versus heliocentric distance, $R$. The grey dashed lines show the $3\sigma$ spread in the fit results (see text). The dashed-dotted red line is a power law fit to the maximum magnetic field in the CME magnetic obstacle $\max(B_{MO}(R))$ versus $R$. The resulting power law formulas are indicated on the plot. (b) Similar plot in log10-log10 space with a linear fit restricted to events $< 1.02$~au. (c) Plot in log10-log10 space and fit for all events (0.07 to 5.4 au).} 
\label{fig:mo_br}
\end{figure*}

\subsection{Magnetic field evolution}

We focus on the evolution of the magnetic field strength within the magnetic obstacle (MO), as this parameter is crucial for empirical CME models describing the global expansion and evolution of CME flux ropes, which contain the strongest magnetic fields.

Previous studies have found different power laws over varying heliocentric distance ranges, in similar formats to Equation 1 (see below), with global expansion power law exponents lower within 1~au, and higher beyond 1~au. For example, \citet{gulisano2010} considered the mean magnetic field strengths within ICMEs using Helios 1 and 2 observations to obtain an exponent of $k=-1.85 \pm 0.07$, covering a heliocentric distance range of 0.3--1~au. This is a much steeper expansion rate than that found by studies using events observed by Ulysses and Juno, covering a heliocentric distance range of 1--5.4~au, with exponents of $k=-1.29 \pm 0.12$ \citep{ebert2009bulk}, $k=-1.21 \pm 0.09$ \citep{richardson2014identification}, and $k=1.24 \pm 0.43$ \citep{davies2021_juno}. 

By combining different spacecraft datasets of ICMEs, previous power law exponents over wider heliocentric distance ranges can be determined. For example, catalogs of ICMEs observed by Ulysses extended their power law relationships between 0.3--5.4 au: \citet{richardson2014identification} included 103 ICMEs observed by the Helios 1 and 2 spacecraft in addition to ICMEs identified at ACE near 1~au in the \citet{richardson2010near} catalog to find $k=-1.38\pm0.03$, \citet{liu2005statistical} found $k=-1.40\pm0.08$ by including ICMEs identified at Helios, Wind, and ACE, and \citet{wang2005characteristics} found $k=-1.52$ by including ICMEs identified at Helios, the Pioneer Venus Orbiter and ACE. Similarly, \cite{davies2021_juno} combined Juno ICME observations with a previous version of the ICMECAT including entries observed by MESSENGER, Venus Express, STEREO-A, STEREO-B, Wind, and MAVEN to determine a power law exponent of $k=-1.66\pm 0.04$ for the flux rope mean magnetic field over the full 0.3--5.4~au heliocentric distance range.

More recently, \cite{salman_2024_psp_cat} analyzed PSP ICME events between 0.23--0.83~au observed up to mid 2022, before most of the ICME observations of PSP very close to the Sun were made. They find a $k=-1.21$ for the mean magnetic obstacle, which is lower than previous studies for the inner heliosphere \citep{gulisano2010,good2019,Salman_2020}. Our aim in this study is now to investigate whether a single power law can accommodate the full distance range from 0.07 to 5.4~au covered in the ICMECAT and whether a connection to solar magnetic field strengths is possible.

We derive power-law fits for both the mean and maximum magnetic field strengths in the magnetic obstacle, $B_{MO}(R)$, which have the form

\begin{equation}
B_{MO}(R)~\text{[nT]}=B_0 \times R~\text{[au]}^{k}, 
\end{equation} 

\noindent where $R$ is the heliocentric distance, and $B_0$ is the power law constant, in our case the magnetic field strength at $R=1$~au. We perform the fitting using the python function \textit{scipy.optimize.curve\underline{\space}fit}. Several algorithms were tested and yielded consistent results. The final results were calculated using the Levenberg-Marquardt algorithm, which provides estimates of the $3\sigma$ spread in the fit parameters using the covariance matrix. These errors should be understood as an uncertainty in the overall power law description, while individual events can differ strongly from the range given within the error bars \citep[e.g.][]{Salman_2020,davies2022}.  Figure~\ref{fig:mo_br}a shows the mean magnetic field in the magnetic obstacle $\langle B_{MO}(R)\rangle$ as a function of heliocentric distance $R$. The resulting power law fit is: 

\begin{equation}
\langle B_{MO}(R)\rangle~\text{[nT]}= (10.72 \pm 0.58) \times R~\text{[au]}^{-1.57 \pm 0.02}. 
\end{equation}

The 6 ICME events observed by PSP close the previously unexplored gap for heliocentric distances $< 0.23$~au that existed prior to this mission, allowing us to extend the power laws to smaller distances down to 0.07~au. 

To test its robustness, we performed the power-law fitting using different radial intervals, from the full range of 0.07--5.4~au down to only 0.07~au--0.5~au, and excluded Ulysses events to avoid biases from high-latitude observations. Across all these variations, the exponent remains stable within $k=[-1.56, -1.58]$. This consistency indicates that the power law is strikingly still applicable to this distance range, yielding no evidence for a change in power-law scaling of the MO magnetic field with distance, even when starting very close to the Sun.

Similarly, we have fitted the same power law to the maximum field strength in the magnetic obstacle, resulting in

\begin{equation}
\max(B_{MO}(R)) \text{[nT]}=(14.92 \pm 0.87)  \times R~\text{[au]}^{-1.53 \pm 0.03},
\end{equation} 

\noindent for which we also find only a small variation in the exponent compared to the previous result for the mean MO fields, with $k=-1.53$ in this case for the maximum field in the MO.

In order to determine whether the innermost ICME observations by PSP ($<0.23$~au) are dominantly influencing the results of the power law fits, we repeat the fitting analysis in log-log space with a linear fit. This makes the range of the dataset smaller so the largest values will not influence the fit results as much, given the steep rise to higher magnitudes close to the Sun. The result is shown in Figure~\ref{fig:mo_br}b for all events $< 1.02$~au (1524 events). There is only a minor change in the power law constant $B_0$, but the power law exponent stays almost the same at $k=-1.58$. This means that for the inner heliosphere ($< 1$~au) this power law exponent is stable, regardless of the method. Extending the log-log fit to the entire heliocentric distance, as shown in Figure~\ref{fig:mo_br}c, results in a slightly different exponent of $k=-1.47$. This means that there is indeed a slight tendency for the PSP observations close to the Sun to dominate the fit results when using a wider distance range, as including events $> 1$~au makes $k$ smaller compared to the fit in linear space ($k=-1.58$). This is consistent with a well-known weaker $k$ in the outer heliosphere \citep[e.g.][]{gulisano2010, davies2021_juno}. Indeed, when we create a fit in log-log space to all events $> 1$~au (448 events) we find $k=-1.32 \pm 0.04$. 

The results for all $k$ parameters for different methods and distance ranges are summarized in Table \ref{tab:fits}. In summary, for practical applications, our analysis results in a scaling law exponent for the mean MO magnetic field in the inner heliosphere ($< 1$~au) as $k=-1.57 \pm 0.03$, and as $k=-1.33 \pm 0.22$ for the outer heliosphere ($> 1$~au).   

\begin{table}[h]
    \centering
    \caption{Fit results for the exponent $k$ using different methods and heliocentric distance domains for the mean total magnetic field in the magnetic obstacle.}
    \begin{tabular}{cccc}
        \hline
        Distance range in au & $k$ linear fit & $k$ log-log fit & number of events\\
        \hline \hline
         0.07--1.02 &  $-1.57 \pm 0.03$  & $-1.58 \pm 0.03$ & 1524 \\
         0.07--5.42 &  $-1.57 \pm 0.02$ & $-1.47 \pm 0.02$ & 1972 \\         
         1.02--5.42 &  $-1.33 \pm 0.22$  & $-1.32 \pm 0.04$ & 448\\
        \hline
    \end{tabular}
    \label{tab:fits}
\end{table}

\subsection{Relation to solar magnetic field strengths}

\begin{figure*}
\centering
{\includegraphics[width=\textwidth]{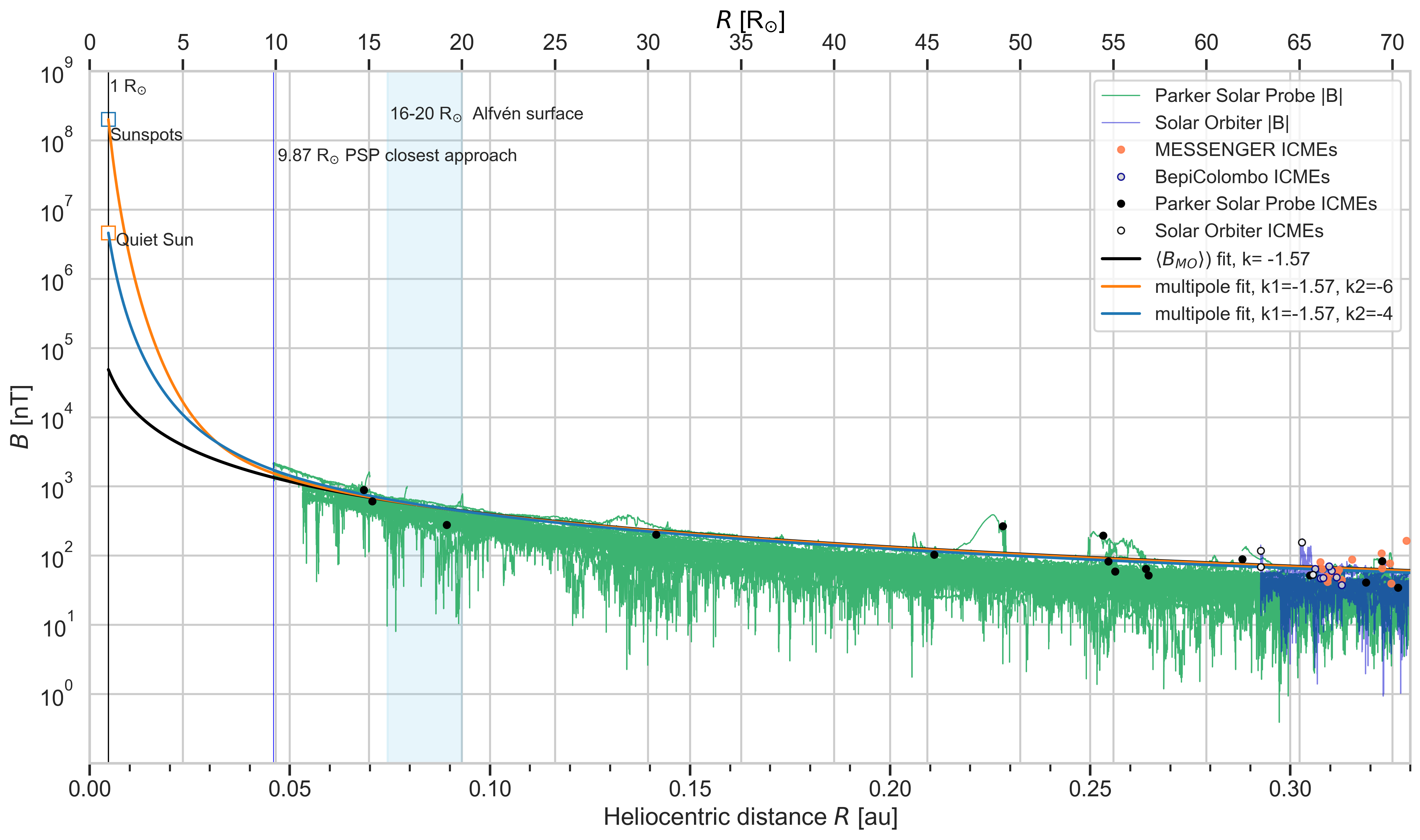}}\caption{Connecting the power law for the ICME magnetic obstacle (MO) to solar magnetic field strengths. The power law for the MO mean magnetic field with an exponent of $k=-1.57$ (solid black line) is compared to continuous in situ observations of the total magnetic field by PSP/FIELDS (thin green line, all data between 2018 October 2 and 2025 April 30) and Solar Orbiter/MAG (thin blue line, 2020 April 15 to 2025 May 31 May). The mean magnetic field in the MO is indicated for ICME events in the ICMECAT by MESSENGER, BepiColombo, PSP and Solar Orbiter as dots (see legend). The multipole power law fit for active regions is represented by the orange solid line, with $k_1=-1.57$ and $k_2=-6$. The multipole power law for the quiet Sun is given by the blue solid line ($k_1=-1.57$ and $k_2=-4$). The black, blue and orange solid lines all merge into a single line around 0.05~au.} 
\label{fig:5_sun_connection}
\end{figure*}


With a stable power law being able to describe the average CME magnetic field evolution over a wide heliocentric distance range, from 0.07 to 5.42~au, we now examine how the field strengths connect to solar observations, given that the smallest distance of PSP during an observation of a magnetic obstacle was only 0.0685~au (or 14.72~$R_{\odot}$) for the 2022 September 5 CME. For context, this distance lies approximately within the same range as the Alfvén surface as determined from previous PSP solar approaches \citep[16--20~$R_{\odot}$,][]{kasper2021alfven}.

Figure~\ref{fig:5_sun_connection} shows the total magnetic field magnitude observed by PSP/FIELDS (October 2018 to 30 April 2025) and Solar Orbiter/MAG (April 2020 to 31 May 2025) throughout the mission durations so far, and we indicate the mean magnetic field of the ICME MOs observed within this distance range as colored dots corresponding to the observing spacecraft. Extending the power law for the mean MO magnetic fields to $1~R_{\odot}$ gives a value of about $5\times 10^4$~nT, or 0.50~Gauss (1~Gauss~$= 10^5$~nT). From studies of the quiet Sun, these average values are known to be around 46~Gauss, or $4.6\times 10^6$~nT \citep[e.g.][]{trellesarjona021}, i.e. 2 magnitudes higher than implied by the ICME MO power law. For active regions, from which many CMEs originate, the discrepancy is even larger: The average magnetic field strengths of sunspots are on the order of kiloGauss, with averages of 2000~Gauss or $2 \times 10^8$~nT \citep[e.g.][]{livingston2006_sunspot}. This demonstrates that a single power law cannot simultaneously describe both near-Sun and interplanetary magnetic field strengths relevant for ICMEs.



In order to find a power law that combines the field strengths for active regions and ICMEs, we extend the power law by adding a second term, reminiscent of a multipole expansion. 
For active regions, we assume an average magnetic field of 2000~Gauss as a representative magnetic field strength at 1 solar radius ($R_{\odot}$).
Hereby we describe this multipole power law with a form of the type

\begin{equation}
B(R) = B_0 \times R^{k_1} + B_1 \times R^{k_2} .
\end{equation}

The first parameter $k_1=-1.57$ is set to match the exponent derived in the previous analysis, and the second exponent $k_2$ is constrained by the ability to connect both the solar and the in situ observations. Here we chose $k_2=-6$, but we note that lower $k_2$ values are also possible to produce a successfull fit and that this choice is not unique. Smaller values for $k_2$ move the distance at which the term with $k_2$ dominates, i.e. the point where the steep ascent begins, closer to the Sun. The power law constants $B_0$ and $B_1$ are then determined using a Levenberg-Marquardt algorithm within \textit{scipy.optimize.curve\_fit} to the non-logarithmic ICMECAT mean MO field values, resulting in:

\begin{equation}
\langle B_{MO}(R)\rangle~\text{[nT]}=10.55 \times R~\text{[au]}^{-1.57} + 2.02\times10^{-6} \times R~\text{[au]}^{-6}. 
\end{equation}

This fit is shown as the orange solid line in Figure \ref{fig:5_sun_connection}.

While this form is not unique, as different $k_2$ values would result in different fit parameters, it nevertheless demonstrates that a multipole power law can link magnetic fields in solar active regions to in situ observations across the heliosphere. However, this approach implies that a sharp increase in field strength must occur at distances $<0.07$~au. A more detailed investigation of the region between 1~$R_{\odot}$ and the first PSP in situ observations starting at 10~$R_{\odot}$ will be necessary to better constrain the magnetic field evolution in this regime, to determine the form of such a power law close to the Sun.



We repeated the same analysis, this time requiring the power law to match the average quiet Sun magnetic field strength of 46~Gauss (solid blue line in Figure \ref{fig:5_sun_connection}). This results in 

\begin{equation}
\langle B_{MO}(R)\rangle~\text{[nT]}=9.92 \times R~\text{[au]}^{-1.57} + 2.1\times10^{-3} \times R~\text{[au]}^{-4}. 
\end{equation}

Here we choose a $k_2=-4$ as the fall-off from the quiet Sun field strengths is expected to be less steep than for active regions, but as before, $k_2$ can be chosen in a small range that would need to be constrained with different means.

\section{Summary and Conclusions} \label{sec:conclusion}

In order to study CME evolution from the Sun to the heliosphere, we have introduced a major update to the ICMECAT living catalog in which we identify ICMEs at five currently active spacecraft, Solar Orbiter, PSP, BepiColombo, STEREO-A, Wind, and 6 other historic missions. This update significantly enhances the coverage of ICME measurements in the inner heliosphere, making the ICMECAT a highly useful resource for the heliophysics research community.


The criteria for an ICME event to be included in the catalog requires a magnetic obstacle (MO) to be present. This choice is motivated by the goal of improving our understanding of the 3D magnetic structure of CME magnetic obstacles. As of today, this is directly only accessible via the 1-D in situ measurements of spacecraft through the enormous CME flux rope structures. MOs often display signatures of flux ropes, featuring smooth rotations of the magnetic field, leading to the strongest geomagnetic storms \citep{zhang2007}. While individual boundaries of the MO by different human observers can vary considerably, they are unlikely to influence the results presented here on general CME evolution.

The catalog now includes 12 ICMEs observed by PSP at distances $< 0.29$~au for the first time. These events were observed by PSP/FIELDS and SWEAP between 2018 and April 2025. With these observations we extend the work of \cite{salman_2024_psp_cat} to even lower heliocentric distances. Using this expanded dataset from 0.07 to 5.4 au, we show that the magnetic field decay of interplanetary coronal mass ejections and their magnetic obstacles can be described well with a power law with exponent $k$ ranging from $k=-1.47$ to $k=-1.57$, depending on the fitting method and the parameter (i.e. mean or maximum magnetic field strength) the power law fit was applied to. We emphasize that our catalog contains only ICMEs with clear magnetic obstacles, and therefore, the results are applicable to this subset of ICME structures.

However, extending this power law to coronal and photospheric distances shows a clear underestimation of these magnetic field strengths of several magnitudes, which we propose to address by introducing multipole power laws with exponents $k_1=-1.57$ and $k_2=-6$ (for photospheric active region field strengths) and $k_2=-4$ (for photospheric quiet Sun field strengths). In comparison, from theory, quadrupole magnetic fields would fall off with distance as $k=-4$. This novel technique should be understood as a first heuristic approach to illustrate the possibility for deriving physically predictive models with such a parameterization in the future. The current direct comparison between photospheric magnetic field strengths and fields in magnetic obstacles is certainly a limitation. To this end, a closer look into how the (modeled) coronal magnetic fields in CME source regions decrease with distance would be very worthwhile, in particular for individual ICME events that are observed by PSP close to the Sun. We leave a closer inspection for future studies. 

Our results provide a statistical description of CME magnetic field evolution, but we emphasize that individual events can deviate significantly in their field decay exponents, for instance due to interaction with high speed streams or other ICMEs \citep[e.g.][]{Davies_2021_solo, salman_2024_psp_cat}. Multi-point imaging and in situ observations of the same CME event \citep[e.g.][]{Davies_2021_solo,palmerio2024mesoscale,palmerio2025event4} are able to give a more detailed picture, and while the numbers of observed events are still relatively low, they are continuously growing with the fleet of spacecraft currently operating.

Power laws as presented here are useful for a wide range of space weather applications, from Faraday rotation methods \citep[e.g.][]{kooi2022_faraday} to flux rope models \citep[e.g.][]{isavnin_2016,weiss_2021_fit,weiss2024distorted, sarkar2020infros,sarkar2024}. For studying solar energetic particle (SEP) propagation \citep[e.g.][]{rodriguez-garcia_2025sep_cloud, wijsen2025}, results of flux rope models often influence how SEPs are accelerated, transported and distributed in the ambient solar wind. It could be possible that remote sensing methods, such as combining observations and models of type IV radio bursts \citep[e.g.][]{husidic2025_gyrotype4}, may shed light on the magnetic field strengths at distances below our in situ limit of 0.07 au, to improve our statistics close to the Sun.

Our results prove particularly useful for ICME magnetic field modeling of in situ observations of ICMEs that are detected near the Sun--Earth line. If a spacecraft far upstream of Earth detects an ICME (at sub-L1 distances), it can transmit the observed in situ magnetic field data to Earth, which results in longer forecast lead times compared to those possible with spacecraft impacted by the ICME at the L1 point. The application of power laws to the sub-L1 magnetic field data then allows to model the field magnitude of the ICME at the distance of Earth, which is useful for improving CME space weather forecasts. This has recently been demonstrated with STEREO-A and Solar Orbiter \citep{laker2024,weiler2025, davies2026_solo}, with first attempts going back to \cite{lindsay1999} and \cite{kubicka2016}. Mission concepts that use spacecraft on distant retrograde orbits \citep[DRO,][]{henon1969, perozzi2017}, which allow spacecraft as part of a fleet to temporarily reside near the Sun--Earth line, so far include ESA HENON, planned to be launched in 2026 \citep{cicalo2025henon}, and ESA SHIELD (as of 2025 in an early planning phase). They build on the Space Weather Diamond \citep{cyr2000} and MIIST mission concepts \citep{lugaz2024ssr}. Further studies need to be undertaken to better understand the CME magnetic field evolution in detail with multi-spacecraft observations from 0.8 to 1.0~au, as these are the distances spacecraft on DROs can be placed on to find an optimal trade-off between forecast lead time, accuracy and number of spacecraft needed for such an orbital configuration.


\section*{Data availability statement}

The ICMECAT living catalog \citep{dataset_icmecatv24} is available online for the research community at \url{https://helioforecast.space/icmecat} and on the filesharing platform figshare as\dataset[doi: 10.6084/m9.figshare.6356420]{https://doi.org/10.6084/m9.figshare.6356420}. For this study, the ICMECAT v2.3, updated on 2025 October 15, version 24 on figshare, has been used. The python code for producing the figures and results for this paper is available on\dataset[github]{https://github.com/cmoestl/icmecat_paper_2026} as release v1.1, and in this figshare repository\dataset[doi: 10.6084/m9.figshare.31421567]{https://doi.org/10.6084/m9.figshare.31421567}.

\begin{acknowledgments}
This work is supported by ERC grant (HELIO4CAST, 10.3030/101042188). Funded by the European Union. Views and opinions expressed are however those of the author(s) only and do not necessarily reflect those of the European Union or the European Research Council Executive Agency. Neither the European Union nor the granting authority can be held responsible for them. Solar Orbiter magnetometer operations are funded by the UK Space Agency (grant UKRI943). T.H. is supported by STFC grant ST/W001071/1. The PSP/FIELDS experiment was developed and is operated under NASA contract NNN06AA01C. D.H. was supported by the German Ministerium für Wirtschaft und Klimaschutz and the German Zentrum für Luft‐ und Raumfahrt under contract 50QW2202.
\end{acknowledgments}

\software{numpy \citep{harris2020numpy}, pandas \citep{mckinney2010pandas, reback2020pandas}, spacepy \citep{niehof2022spacepy}, astropy \citep{astropy2013, astropy2018, astropy2022}, spiceypy \citep{annex2020spiceypy}, scipy \citep{virtanen2020scipy}, plotly \citep{plotly2026} and matplotlib \citep{hunter2007matplotlib}.}

\bibliography{chrisbib}{}

@ARTICLE{yogesh2026,
       author = {{Yogesh} and {Ofman}, Leon and {Klein}, Kristopher and {Shankarappa}, Niranjana and {Martinovi{\'c}}, Mihailo M. and {Howes}, Gregory G. and {Mostafavi}, Parisa and {Boardsen}, Scott A and {Sadykov}, Viacheslav M and {Pal}, Sanchita and {Jian}, Lan K and {Gupta}, Aakash and {Chakrabarty}, D. and {Alterman}, B.~L. and {Verniero}, Jaye L and {Paulson}, K.~W. and {Huang}, Jia and {Livi}, Roberto and {Larson}, Davin E. and {M{\"o}stl}, Christian and {Davies}, Emma E. and {Weiler}, Eva},
        title = "{Solar Wind Heating Near the Sun: A Radial Evolution Approach}",
      journal = {arXiv e-prints},
         year = 2026,
        month = feb,
          eid = {arXiv:2602.10275},
        pages = {arXiv:2602.10275},
          doi = {10.48550/arXiv.2602.10275},
archivePrefix = {arXiv},
       eprint = {2602.10275},
 primaryClass = {astro-ph.SR},
       adsurl = {https://ui.adsabs.harvard.edu/abs/2026arXiv260210275Y}
}

@Article{harris2020numpy,
 title         = {Array programming with {NumPy}},
 author        = {Charles R. Harris and K. Jarrod Millman and St{\'{e}}fan J.
                 van der Walt and Ralf Gommers and Pauli Virtanen and David
                 Cournapeau and Eric Wieser and Julian Taylor and Sebastian
                 Berg and Nathaniel J. Smith and Robert Kern and Matti Picus
                 and Stephan Hoyer and Marten H. van Kerkwijk and Matthew
                 Brett and Allan Haldane and Jaime Fern{\'{a}}ndez del
                 R{\'{i}}o and Mark Wiebe and Pearu Peterson and Pierre
                 G{\'{e}}rard-Marchant and Kevin Sheppard and Tyler Reddy and
                 Warren Weckesser and Hameer Abbasi and Christoph Gohlke and
                 Travis E. Oliphant},
 year          = {2020},
 month         = sep,
 journal       = {Nature},
 volume        = {585},
 number        = {7825},
 pages         = {357--362},
 doi           = {10.1038/s41586-020-2649-2},
 publisher     = {Springer Science and Business Media {LLC}},
 url           = {https://doi.org/10.1038/s41586-020-2649-2}
}

@article{astropy2013,
Adsurl = {http://adsabs.harvard.edu/abs/2013A%26A...558A..33A},
Archiveprefix = {arXiv},
Author = {{Astropy Collaboration} and {Robitaille}, T.~P. and {Tollerud}, E.~J. and {Greenfield}, P. and {Droettboom}, M. and {Bray}, E. and {Aldcroft}, T. and {Davis}, M. and {Ginsburg}, A. and {Price-Whelan}, A.~M. and {Kerzendorf}, W.~E. and {Conley}, A. and {Crighton}, N. and {Barbary}, K. and {Muna}, D. and {Ferguson}, H. and {Grollier}, F. and {Parikh}, M.~M. and {Nair}, P.~H. and {Unther}, H.~M. and {Deil}, C. and {Woillez}, J. and {Conseil}, S. and {Kramer}, R. and {Turner}, J.~E.~H. and {Singer}, L. and {Fox}, R. and {Weaver}, B.~A. and {Zabalza}, V. and {Edwards}, Z.~I. and {Azalee Bostroem}, K. and {Burke}, D.~J. and {Casey}, A.~R. and {Crawford}, S.~M. and {Dencheva}, N. and {Ely}, J. and {Jenness}, T. and {Labrie}, K. and {Lim}, P.~L. and {Pierfederici}, F. and {Pontzen}, A. and {Ptak}, A. and {Refsdal}, B. and {Servillat}, M. and {Streicher}, O.},
Doi = {10.1051/0004-6361/201322068},
Eid = {A33},
Eprint = {1307.6212},
Journal = {\aap},
Month = oct,
Pages = {A33},
Primaryclass = {astro-ph.IM},
Title = {{Astropy: A community Python package for astronomy}},
Volume = 558,
Year = 2013}

@ARTICLE{astropy2018,
       author = {{Astropy Collaboration} and {Price-Whelan}, A.~M. and
         {Sip{\H{o}}cz}, B.~M. and {G{\"u}nther}, H.~M. and {Lim}, P.~L. and
         {Crawford}, S.~M. and {Conseil}, S. and {Shupe}, D.~L. and
         {Craig}, M.~W. and {Dencheva}, N. and {Ginsburg}, A. and {Vand
        erPlas}, J.~T. and {Bradley}, L.~D. and {P{\'e}rez-Su{\'a}rez}, D. and
         {de Val-Borro}, M. and {Aldcroft}, T.~L. and {Cruz}, K.~L. and
         {Robitaille}, T.~P. and {Tollerud}, E.~J. and {Ardelean}, C. and
         {Babej}, T. and {Bach}, Y.~P. and {Bachetti}, M. and {Bakanov}, A.~V. and
         {Bamford}, S.~P. and {Barentsen}, G. and {Barmby}, P. and
         {Baumbach}, A. and {Berry}, K.~L. and {Biscani}, F. and {Boquien}, M. and
         {Bostroem}, K.~A. and {Bouma}, L.~G. and {Brammer}, G.~B. and
         {Bray}, E.~M. and {Breytenbach}, H. and {Buddelmeijer}, H. and
         {Burke}, D.~J. and {Calderone}, G. and {Cano Rodr{\'\i}guez}, J.~L. and
         {Cara}, M. and {Cardoso}, J.~V.~M. and {Cheedella}, S. and {Copin}, Y. and
         {Corrales}, L. and {Crichton}, D. and {D'Avella}, D. and {Deil}, C. and
         {Depagne}, {\'E}. and {Dietrich}, J.~P. and {Donath}, A. and
         {Droettboom}, M. and {Earl}, N. and {Erben}, T. and {Fabbro}, S. and
         {Ferreira}, L.~A. and {Finethy}, T. and {Fox}, R.~T. and
         {Garrison}, L.~H. and {Gibbons}, S.~L.~J. and {Goldstein}, D.~A. and
         {Gommers}, R. and {Greco}, J.~P. and {Greenfield}, P. and
         {Groener}, A.~M. and {Grollier}, F. and {Hagen}, A. and {Hirst}, P. and
         {Homeier}, D. and {Horton}, A.~J. and {Hosseinzadeh}, G. and {Hu}, L. and
         {Hunkeler}, J.~S. and {Ivezi{\'c}}, {\v{Z}}. and {Jain}, A. and
         {Jenness}, T. and {Kanarek}, G. and {Kendrew}, S. and {Kern}, N.~S. and
         {Kerzendorf}, W.~E. and {Khvalko}, A. and {King}, J. and {Kirkby}, D. and
         {Kulkarni}, A.~M. and {Kumar}, A. and {Lee}, A. and {Lenz}, D. and
         {Littlefair}, S.~P. and {Ma}, Z. and {Macleod}, D.~M. and
         {Mastropietro}, M. and {McCully}, C. and {Montagnac}, S. and
         {Morris}, B.~M. and {Mueller}, M. and {Mumford}, S.~J. and {Muna}, D. and
         {Murphy}, N.~A. and {Nelson}, S. and {Nguyen}, G.~H. and
         {Ninan}, J.~P. and {N{\"o}the}, M. and {Ogaz}, S. and {Oh}, S. and
         {Parejko}, J.~K. and {Parley}, N. and {Pascual}, S. and {Patil}, R. and
         {Patil}, A.~A. and {Plunkett}, A.~L. and {Prochaska}, J.~X. and
         {Rastogi}, T. and {Reddy Janga}, V. and {Sabater}, J. and
         {Sakurikar}, P. and {Seifert}, M. and {Sherbert}, L.~E. and
         {Sherwood-Taylor}, H. and {Shih}, A.~Y. and {Sick}, J. and
         {Silbiger}, M.~T. and {Singanamalla}, S. and {Singer}, L.~P. and
         {Sladen}, P.~H. and {Sooley}, K.~A. and {Sornarajah}, S. and
         {Streicher}, O. and {Teuben}, P. and {Thomas}, S.~W. and
         {Tremblay}, G.~R. and {Turner}, J.~E.~H. and {Terr{\'o}n}, V. and
         {van Kerkwijk}, M.~H. and {de la Vega}, A. and {Watkins}, L.~L. and
         {Weaver}, B.~A. and {Whitmore}, J.~B. and {Woillez}, J. and
         {Zabalza}, V. and {Astropy Contributors}},
        title = "{The Astropy Project: Building an Open-science Project and Status of the v2.0 Core Package}",
      journal = {\aj},
         year = 2018,
        month = sep,
       volume = {156},
       number = {3},
          eid = {123},
        pages = {123},
          doi = {10.3847/1538-3881/aabc4f},
archivePrefix = {arXiv},
       eprint = {1801.02634},
 primaryClass = {astro-ph.IM},
       adsurl = {https://ui.adsabs.harvard.edu/abs/2018AJ....156..123A}
}

@ARTICLE{astropy2022,
       author = {{Astropy Collaboration} and {Price-Whelan}, Adrian M. and {Lim}, Pey Lian and {Earl}, Nicholas and {Starkman}, Nathaniel and {Bradley}, Larry and {Shupe}, David L. and {Patil}, Aarya A. and {Corrales}, Lia and {Brasseur}, C.~E. and {N{"o}the}, Maximilian and {Donath}, Axel and {Tollerud}, Erik and {Morris}, Brett M. and {Ginsburg}, Adam and {Vaher}, Eero and {Weaver}, Benjamin A. and {Tocknell}, James and {Jamieson}, William and {van Kerkwijk}, Marten H. and {Robitaille}, Thomas P. and {Merry}, Bruce and {Bachetti}, Matteo and {G{"u}nther}, H. Moritz and {Aldcroft}, Thomas L. and {Alvarado-Montes}, Jaime A. and {Archibald}, Anne M. and {B{'o}di}, Attila and {Bapat}, Shreyas and {Barentsen}, Geert and {Baz{'a}n}, Juanjo and {Biswas}, Manish and {Boquien}, M{'e}d{'e}ric and {Burke}, D.~J. and {Cara}, Daria and {Cara}, Mihai and {Conroy}, Kyle E. and {Conseil}, Simon and {Craig}, Matthew W. and {Cross}, Robert M. and {Cruz}, Kelle L. and {D'Eugenio}, Francesco and {Dencheva}, Nadia and {Devillepoix}, Hadrien A.~R. and {Dietrich}, J{"o}rg P. and {Eigenbrot}, Arthur Davis and {Erben}, Thomas and {Ferreira}, Leonardo and {Foreman-Mackey}, Daniel and {Fox}, Ryan and {Freij}, Nabil and {Garg}, Suyog and {Geda}, Robel and {Glattly}, Lauren and {Gondhalekar}, Yash and {Gordon}, Karl D. and {Grant}, David and {Greenfield}, Perry and {Groener}, Austen M. and {Guest}, Steve and {Gurovich}, Sebastian and {Handberg}, Rasmus and {Hart}, Akeem and {Hatfield-Dodds}, Zac and {Homeier}, Derek and {Hosseinzadeh}, Griffin and {Jenness}, Tim and {Jones}, Craig K. and {Joseph}, Prajwel and {Kalmbach}, J. Bryce and {Karamehmetoglu}, Emir and {Ka{l}uszy{'n}ski}, Miko{l}aj and {Kelley}, Michael S.~P. and {Kern}, Nicholas and {Kerzendorf}, Wolfgang E. and {Koch}, Eric W. and {Kulumani}, Shankar and {Lee}, Antony and {Ly}, Chun and {Ma}, Zhiyuan and {MacBride}, Conor and {Maljaars}, Jakob M. and {Muna}, Demitri and {Murphy}, N.~A. and {Norman}, Henrik and {O'Steen}, Richard and {Oman}, Kyle A. and {Pacifici}, Camilla and {Pascual}, Sergio and {Pascual-Granado}, J. and {Patil}, Rohit R. and {Perren}, Gabriel I. and {Pickering}, Timothy E. and {Rastogi}, Tanuj and {Roulston}, Benjamin R. and {Ryan}, Daniel F. and {Rykoff}, Eli S. and {Sabater}, Jose and {Sakurikar}, Parikshit and {Salgado}, Jes{'u}s and {Sanghi}, Aniket and {Saunders}, Nicholas and {Savchenko}, Volodymyr and {Schwardt}, Ludwig and {Seifert-Eckert}, Michael and {Shih}, Albert Y. and {Jain}, Anany Shrey and {Shukla}, Gyanendra and {Sick}, Jonathan and {Simpson}, Chris and {Singanamalla}, Sudheesh and {Singer}, Leo P. and {Singhal}, Jaladh and {Sinha}, Manodeep and {Sip{H{o}}cz}, Brigitta M. and {Spitler}, Lee R. and {Stansby}, David and {Streicher}, Ole and {{{S}}umak}, Jani and {Swinbank}, John D. and {Taranu}, Dan S. and {Tewary}, Nikita and {Tremblay}, Grant R. and {Val-Borro}, Miguel de and {Van Kooten}, Samuel J. and {Vasovi{'c}}, Zlatan and {Verma}, Shresth and {de Miranda Cardoso}, Jos{'e} Vin{'i}cius and {Williams}, Peter K.~G. and {Wilson}, Tom J. and {Winkel}, Benjamin and {Wood-Vasey}, W.~M. and {Xue}, Rui and {Yoachim}, Peter and {Zhang}, Chen and {Zonca}, Andrea and {Astropy Project Contributors}},
        title = "{The Astropy Project: Sustaining and Growing a Community-oriented Open-source Project and the Latest Major Release (v5.0) of the Core Package}",
      journal = {\apj},
         year = 2022,
        month = aug,
       volume = {935},
       number = {2},
          eid = {167},
        pages = {167},
          doi = {10.3847/1538-4357/ac7c74},
archivePrefix = {arXiv},
       eprint = {2206.14220},
 primaryClass = {astro-ph.IM},
       adsurl = {https://ui.adsabs.harvard.edu/abs/2022ApJ...935..167A}
}

@Article{hunter2007matplotlib,
  Author    = {Hunter, J. D.},
  Title     = {Matplotlib: A 2D graphics environment},
  Journal   = {Computing in Science \& Engineering},
  Volume    = {9},
  Number    = {3},
  Pages     = {90--95},
  publisher = {IEEE COMPUTER SOC},
  doi       = {10.1109/MCSE.2007.55},
  year      = 2007
}

@article{niehof2022spacepy, 
title={The SpacePy space science package at 12 years}, 
author={Niehof, Jonathan T and Morley, Steven K and Welling, Daniel T and Larsen, Brian A}, 
journal={Frontiers in Astronomy and Space Sciences}, 
volume={9}, 
year={2022}, 
doi={10.3389/fspas.2022.1023612}, 
publisher={Frontiers} 
}

@misc{plotly2026,
title={An interactive, open-source, and browser-based graphing library for Python},
author = {Kruchten, Nicolas and Seier, Andrew and Parmer, Chris},
year={2026},
doi={10.5281/zenodo.14503524},
url={"https://github.com/plotly/plotly.py"}
}

@software{reback2020pandas,
    author       = {{The Pandas Development Team}},
    title        = {pandas-dev/pandas: Pandas},
    month        = feb,
    year         = 2020,
    publisher    = {Zenodo},
    version      = {latest},
    doi          = {10.5281/zenodo.3509134},
    url          = {https://doi.org/10.5281/zenodo.3509134}
}

@InProceedings{mckinney2010pandas,
  author    = {{M}c{K}inney, {W}es},
  title     = {{D}ata {S}tructures for {S}tatistical {C}omputing in {P}ython},
  booktitle = {{P}roceedings of the 9th {P}ython in {S}cience {C}onference},
  pages     = {56 - 61},
  year      = {2010},
  editor    = {{S}t\'efan van der {W}alt and {J}arrod {M}illman},
  doi       = {10.25080/Majora-92bf1922-00a}
}

@ARTICLE{annex2020spiceypy,
       author = {{Annex}, Andrew and {Pearson}, Ben and {Seignovert}, Beno{\^\i}t and {Carcich}, Brian and {Eichhorn}, Helge and {Mapel}, Jesse and {von Forstner}, Johan and {McAuliffe}, Jonathan and {del Rio}, Jorge and {Berry}, Kristin and {Aye}, K. -Michael and {Stefko}, Marcel and {de Val-Borro}, Miguel and {Kulumani}, Shankar and {Murakami}, Shin-ya},
        title = "{SpiceyPy: a Pythonic Wrapper for the SPICE Toolkit}",
      journal = {The Journal of Open Source Software},
         year = 2020,
        month = feb,
       volume = {5},
       number = {46},
          eid = {2050},
        pages = {2050},
          doi = {10.21105/joss.02050},
       adsurl = {https://ui.adsabs.harvard.edu/abs/2020JOSS....5.2050A}
}

@ARTICLE{virtanen2020scipy,
       author = {{Virtanen}, Pauli and {Gommers}, Ralf and {Oliphant}, Travis E. and {Haberland}, Matt and {Reddy}, Tyler and {Cournapeau}, David and {Burovski}, Evgeni and {Peterson}, Pearu and {Weckesser}, Warren and {Bright}, Jonathan and {van der Walt}, St{\'e}fan J. and {Brett}, Matthew and {Wilson}, Joshua and {Millman}, K. Jarrod and {Mayorov}, Nikolay and {Nelson}, Andrew R.~J. and {Jones}, Eric and {Kern}, Robert and {Larson}, Eric and {Carey}, C.~J. and {Polat}, {\.I}lhan and {Feng}, Yu and {Moore}, Eric W. and {VanderPlas}, Jake and {Laxalde}, Denis and {Perktold}, Josef and {Cimrman}, Robert and {Henriksen}, Ian and {Quintero}, E.~A. and {Harris}, Charles R. and {Archibald}, Anne M. and {Ribeiro}, Ant{\^o}nio H. and {Pedregosa}, Fabian and {van Mulbregt}, Paul and {SciPy 1. 0 Contributors}},
        title = "{SciPy 1.0: fundamental algorithms for scientific computing in Python}",
      journal = {Nature Methods},
         year = 2020,
        month = feb,
       volume = {17},
        pages = {261-272},
          doi = {10.1038/s41592-019-0686-2},
archivePrefix = {arXiv},
       eprint = {1907.10121},
 primaryClass = {cs.MS},
       adsurl = {https://ui.adsabs.harvard.edu/abs/2020NatMe..17..261V}
}

@ARTICLE{sarkar2020infros,
       author = {{Sarkar}, Ranadeep and {Gopalswamy}, Nat and {Srivastava}, Nandita},
        title = "{An Observationally Constrained Analytical Model for Predicting the Magnetic Field Vectors of Interplanetary Coronal Mass Ejections at 1 au}",
      journal = {\apj},
         year = 2020,
        month = jan,
       volume = {888},
       number = {2},
          eid = {121},
        pages = {121},
          doi = {10.3847/1538-4357/ab5fd7},
archivePrefix = {arXiv},
       eprint = {1912.03494},
 primaryClass = {astro-ph.SR},
       adsurl = {https://ui.adsabs.harvard.edu/abs/2020ApJ...888..121S}
}

@ARTICLE{rodriguez-garcia_2025sep_cloud,
       author = {{Rodr{\'\i}guez-Garc{\'\i}a}, L. and {G{\'o}mez-Herrero}, R. and {Dresing}, N. and {Balmaceda}, L.~A. and {Palmerio}, E. and {Kouloumvakos}, A. and {Jebaraj}, I.~C. and {Espinosa Lara}, F. and {Roco}, M. and {Palmroos}, C. and {Warmuth}, A. and {Nicolaou}, G. and {Mason}, G.~M. and {Guo}, J. and {Laitinen}, T. and {Cernuda}, I. and {Nieves-Chinchilla}, T. and {Fedeli}, A. and {Lee}, C.~O. and {Cohen}, C.~M.~S. and {Owen}, C.~J. and {Ho}, G.~C. and {Malandraki}, O. and {Vainio}, R. and {Rodr{\'\i}guez-Pacheco}, J.},
        title = "{Solar energetic particles injected inside and outside a magnetic cloud: The widespread solar energetic particle event on 2022 January 20}",
      journal = {\aap},
         year = 2025,
        month = feb,
       volume = {694},
          eid = {A64},
        pages = {A64},
          doi = {10.1051/0004-6361/202452158},
archivePrefix = {arXiv},
       eprint = {2409.04564},
 primaryClass = {astro-ph.SR},
       adsurl = {https://ui.adsabs.harvard.edu/abs/2025A&A...694A..64R}
}

@ARTICLE{husidic2025_gyrotype4,
       author = {{Husidic}, E. and {Wijsen}, N. and {Jebaraj}, I.~C. and {Vourlidas}, A. and {Linan}, L. and {Vainio}, R. and {Poedts}, S.},
        title = "{Modelling gyrosynchrotron emission from coronal energetic electrons in a CME flux rope}",
      journal = {\aap},
         year = 2025,
        month = sep,
       volume = {701},
          eid = {A53},
        pages = {A53},
          doi = {10.1051/0004-6361/202555534},
archivePrefix = {arXiv},
       eprint = {2507.16449},
 primaryClass = {astro-ph.SR},
       adsurl = {https://ui.adsabs.harvard.edu/abs/2025A&A...701A..53H}
}

@ARTICLE{wijsen2025,
       author = {{Wijsen}, N. and {Jebaraj}, I.~C. and {Dresing}, N. and {Kouloumvakos}, A. and {Palmerio}, E. and {Rodr{\'\i}guez-Garc{\'\i}a}, L.},
        title = "{Freely propagating flanks of wide coronal-mass-ejection-driven shocks: Modelling and observational insights}",
      journal = {\aap},
         year = 2025,
        month = jul,
       volume = {699},
          eid = {A51},
        pages = {A51},
          doi = {10.1051/0004-6361/202453598},
archivePrefix = {arXiv},
       eprint = {2505.02794},
 primaryClass = {physics.space-ph},
       adsurl = {https://ui.adsabs.harvard.edu/abs/2025A&A...699A..51W}
}

@ARTICLE{palmerio2024mesoscale,
       author = {{Palmerio}, Erika and {Carcaboso}, Fernando and {Khoo}, Leng Ying and {Salman}, Tarik M. and {S{\'a}nchez-Cano}, Beatriz and {Lynch}, Benjamin J. and {Rivera}, Yeimy J. and {Pal}, Sanchita and {Nieves-Chinchilla}, Teresa and {Weiss}, Andreas J. and {Lario}, David and {Mieth}, Johannes Z.~D. and {Heyner}, Daniel and {Stevens}, Michael L. and {Romeo}, Orlando M. and {Zhukov}, Andrei N. and {Rodriguez}, Luciano and {Lee}, Christina O. and {Cohen}, Christina M.~S. and {Rodr{\'\i}guez-Garc{\'\i}a}, Laura and {Whittlesey}, Phyllis L. and {Dresing}, Nina and {Oleynik}, Philipp and {Jebaraj}, Immanuel C. and {Fischer}, David and {Schmid}, Daniel and {Richter}, Ingo and {Auster}, Hans-Ulrich and {Fraschetti}, Federico and {Mierla}, Marilena},
        title = "{On the Mesoscale Structure of Coronal Mass Ejections at Mercury's Orbit: BepiColombo and Parker Solar Probe Observations}",
      journal = {\apj},
         year = 2024,
        month = mar,
       volume = {963},
       number = {2},
          eid = {108},
        pages = {108},
          doi = {10.3847/1538-4357/ad1ab4},
archivePrefix = {arXiv},
       eprint = {2401.01875},
 primaryClass = {astro-ph.SR},
       adsurl = {https://ui.adsabs.harvard.edu/abs/2024ApJ...963..108P}
}

@dataset{dataset_icmecatv24,
doi = {10.6084/M9.FIGSHARE.6356420.V24},
url = {https://figshare.com/articles/dataset/HELCATS_Interplanetary_Coronal_Mass_Ejection_Catalog_v2_0/6356420/24},
author = {Möstl, Christian and Davies, Emma and Weiler, Eva},
title = {HELIO4CAST Interplanetary Coronal Mass Ejection Catalog v2.3},
publisher = {figshare},
year = {2025},
copyright = {MIT License}
}

@ARTICLE{pomoell2019active,
       author = {{Pomoell}, Jens and {Lumme}, Erkka and {Kilpua}, Emilia},
        title = "{Time-dependent Data-driven Modeling of Active Region Evolution Using Energy-optimized Photospheric Electric Fields}",
      journal = {\solphys},
         year = 2019,
        month = apr,
       volume = {294},
       number = {4},
          eid = {41},
        pages = {41},
          doi = {10.1007/s11207-019-1430-x},
       adsurl = {https://ui.adsabs.harvard.edu/abs/2019SoPh..294...41P}
}

@ARTICLE{wang2005characteristics,
       author = {{Wang}, C. and {Du}, D. and {Richardson}, J.~D.},
        title = "{Characteristics of the interplanetary coronal mass ejections in the heliosphere between 0.3 and 5.4 AU}",
      journal = {\jgr (Space Physics)},
         year = 2005,
        month = oct,
       volume = {110},
       number = {A10},
          eid = {A10107},
        pages = {A10107},
          doi = {10.1029/2005JA011198},
       adsurl = {https://ui.adsabs.harvard.edu/abs/2005JGRA..11010107W}
}

@ARTICLE{liu2005statistical,
       author = {{Liu}, Y. and {Richardson}, J.~D. and {Belcher}, J.~W.},
        title = "{A statistical study of the properties of interplanetary coronal mass ejections from 0.3 to 5.4 AU}",
      journal = {\planss},
         year = 2005,
        month = jan,
       volume = {53},
       number = {1-3},
        pages = {3-17},
          doi = {10.1016/j.pss.2004.09.023},
       adsurl = {https://ui.adsabs.harvard.edu/abs/2005P&SS...53....3L}
}

@ARTICLE{richardson2010near,
       author = {{Richardson}, I.~G. and {Cane}, H.~V.},
        title = "{Near-Earth Interplanetary Coronal Mass Ejections During Solar Cycle 23 (1996 - 2009): Catalog and Summary of Properties}",
      journal = {\solphys},
         year = 2010,
        month = jun,
       volume = {264},
       number = {1},
        pages = {189-237},
          doi = {10.1007/s11207-010-9568-6},
       adsurl = {https://ui.adsabs.harvard.edu/abs/2010SoPh..264..189R}
}

@ARTICLE{ebert2009bulk,
       author = {{Ebert}, R.~W. and {McComas}, D.~J. and {Elliott}, H.~A. and {Forsyth}, R.~J. and {Gosling}, J.~T.},
        title = "{Bulk properties of the slow and fast solar wind and interplanetary coronal mass ejections measured by Ulysses: Three polar orbits of observations}",
      journal = {\jgr (Space Physics)},
         year = 2009,
        month = jan,
       volume = {114},
       number = {A1},
          eid = {A01109},
        pages = {A01109},
          doi = {10.1029/2008JA013631},
       adsurl = {https://ui.adsabs.harvard.edu/abs/2009JGRA..114.1109E}
}

@ARTICLE{gulisano2010,
       author = {{Gulisano}, A.~M. and {D{\'e}moulin}, P. and {Dasso}, S. and {Ruiz}, M.~E. and {Marsch}, E.},
        title = "{Global and local expansion of magnetic clouds in the inner heliosphere}",
      journal = {\aap},
         year = 2010,
        month = jan,
       volume = {509},
          eid = {A39},
        pages = {A39},
          doi = {10.1051/0004-6361/200912375},
archivePrefix = {arXiv},
       eprint = {1206.1112},
 primaryClass = {physics.space-ph},
       adsurl = {https://ui.adsabs.harvard.edu/abs/2010A&A...509A..39G}
}

@ARTICLE{paouris2023_sep2022,
       author = {{Paouris}, Evangelos and {Vourlidas}, Angelos and {Kouloumvakos}, Athanasios and {Papaioannou}, Athanasios and {Jagarlamudi}, Vamsee Krishna and {Horbury}, Timothy},
        title = "{The Space Weather Context of the First Extreme Event of Solar Cycle 25, on 2022 September 5}",
      journal = {\apj},
         year = 2023,
        month = oct,
       volume = {956},
       number = {1},
          eid = {58},
        pages = {58},
          doi = {10.3847/1538-4357/acf30f},
       adsurl = {https://ui.adsabs.harvard.edu/abs/2023ApJ...956...58P}
}

@ARTICLE{romeo2023_sep2022,
       author = {{Romeo}, O.~M. and {Braga}, C.~R. and {Badman}, S.~T. and {Larson}, D.~E. and {Stevens}, M.~L. and {Huang}, J. and {Phan}, T. and {Rahmati}, A. and {Livi}, R. and {Alnussirat}, S.~T. and {Whittlesey}, P.~L. and {Szabo}, A. and {Klein}, K.~G. and {Niembro-Hernandez}, T. and {Paulson}, K. and {Verniero}, J.~L. and {Lario}, D. and {Raouafi}, N.~E. and {Ervin}, T. and {Kasper}, J. and {Pulupa}, M. and {Bale}, S.~D. and {Linton}, M.~G.},
        title = "{Near-Sun In Situ and Remote-sensing Observations of a Coronal Mass Ejection and its Effect on the Heliospheric Current Sheet}",
      journal = {\apj},
         year = 2023,
        month = sep,
       volume = {954},
       number = {2},
          eid = {168},
        pages = {168},
          doi = {10.3847/1538-4357/ace62e},
       adsurl = {https://ui.adsabs.harvard.edu/abs/2023ApJ...954..168R}
}

@ARTICLE{riley2025_sep2022,
       author = {{Riley}, Pete and {Ben-Nun}, Michal and {Gonz{\'a}lez-Avil{\'e}s}, J.~J. and {Palmerio}, Erika and {T{\"o}r{\"o}k}, Tibor and {Linker}, Jon A. and {Kouloumvakos}, Athanasios and {Romeo}, Orlando M. and {Ervin}, Tamar and {S{\'a}nchez-Cano}, Beatriz and {Varsani}, Ali and {Laky}, Gunter and {Jeszenszky}, Harald and {Orsini}, Stefano and {Milillo}, Anna and {Heyner}, Daniel and {Auster}, Hans-Ulrich and {Richter}, Ingo and {Schmid}, Daniel and {Fischer}, David},
        title = "{Understanding the global structure of the September 5, 2022, coronal mass ejection using sunRunner3D}",
      journal = {Journal of Space Weather and Space Climate},
         year = 2025,
        month = may,
       volume = {15},
          eid = {17},
        pages = {17},
          doi = {10.1051/swsc/2025010},
       adsurl = {https://ui.adsabs.harvard.edu/abs/2025JSWSC..15...17R}
}

@ARTICLE{jebaraj2024,
       author = {{Jebaraj}, I.~C. and {Agapitov}, O.~V. and {Gedalin}, M. and {Vuorinen}, L. and {Miceli}, M. and {Cohen}, C.~M.~S. and {Voshchepynets}, A. and {Kouloumvakos}, A. and {Dresing}, N. and {Marmyleva}, A. and {Krasnoselskikh}, V. and {Balikhin}, M. and {Mitchell}, J.~G. and {Labrador}, A.~W. and {Wijsen}, N. and {Palmerio}, E. and {Colomban}, L. and {Pomoell}, J. and {Kilpua}, E.~K.~J. and {Pulupa}, M. and {Mozer}, F.~S. and {Raouafi}, N.~E. and {McComas}, D.~J. and {Bale}, S.~D. and {Vainio}, R.},
        title = "{Direct Measurements of Synchrotron-emitting Electrons at Near-Sun Shocks}",
      journal = {\apjl},
         year = 2024,
        month = nov,
       volume = {976},
       number = {1},
          eid = {L7},
        pages = {L7},
          doi = {10.3847/2041-8213/ad8eb8},
archivePrefix = {arXiv},
       eprint = {2410.15933},
 primaryClass = {physics.space-ph},
       adsurl = {https://ui.adsabs.harvard.edu/abs/2024ApJ...976L...7J}
}

@ARTICLE{zhang2025_multipoint,
       author = {{Zhang}, Zhiyong and {Shen}, Chenglong and {Chi}, Yutian and {Mao}, Dongwei and {Liu}, Junyan and {Xu}, Mengjiao and {Zhong}, Zhihui and {Luo}, Jingyu and {Wang}, Can and {Wang}, Yuming},
        title = "{Studying the Evolution of ICMEs in the Heliosphere Through Multipoint Observations}",
      journal = {Journal of Geophysical Research (Space Physics)},
         year = 2025,
        month = jul,
       volume = {130},
       number = {7},
          eid = {e2025JA034094},
        pages = {e2025JA034094},
          doi = {10.1029/2025JA034094},
       adsurl = {https://ui.adsabs.harvard.edu/abs/2025JGRA..13034094Z}
}

@ARTICLE{braga2024_june2022,
       author = {{Braga}, Carlos R. and {Jagarlamudi}, Vamsee Krishna and {Vourlidas}, Angelos and {Stenborg}, Guillermo and {Nieves-Chinchilla}, Teresa},
        title = "{A Coronal Mass Ejection Impacting Parker Solar Probe at 14 Solar Radii}",
      journal = {\apj},
         year = 2024,
        month = apr,
       volume = {965},
       number = {2},
          eid = {185},
        pages = {185},
          doi = {10.3847/1538-4357/ad2b4e},
       adsurl = {https://ui.adsabs.harvard.edu/abs/2024ApJ...965..185B}
}

@ARTICLE{livingston2006_sunspot,
       author = {{Livingston}, W. and {Harvey}, J.~W. and {Malanushenko}, O.~V. and {Webster}, L.},
        title = "{Sunspots with the Strongest Magnetic Fields}",
      journal = {\solphys},
         year = 2006,
        month = dec,
       volume = {239},
       number = {1-2},
        pages = {41-68},
          doi = {10.1007/s11207-006-0265-4},
       adsurl = {https://ui.adsabs.harvard.edu/abs/2006SoPh..239...41L}
}

@ARTICLE{temmer2023,
       author = {{Temmer}, M. and {Scolini}, C. and {Richardson}, I.~G. and {Heinemann}, S.~G. and {Paouris}, E. and {Vourlidas}, A. and {Bisi}, M.~M. and {writing teams} and {:} and {Al-Haddad}, N. and {Amerstorfer}, T. and {Barnard}, L. and {Buresova}, D. and {Hofmeister}, S.~J. and {Iwai}, K. and {Jackson}, B.~V. and {Jarolim}, R. and {Jian}, L.~K. and {Linker}, J.~A. and {Lugaz}, N. and {Manoharan}, P.~K. and {Mays}, M.~L. and {Mishra}, W. and {Owens}, M.~J. and {Palmerio}, E. and {Perri}, B. and {Pomoell}, J. and {Pinto}, R.~F. and {Samara}, E. and {Singh}, T. and {Sur}, D. and {Verbeke}, C. and {Veronig}, A.~M. and {Zhuang}, B.},
        title = "{CME Propagation Through the Heliosphere: Status and Future of Observations and Model Development}",
      journal = {arXiv e-prints},
         year = 2023,
        month = aug,
          eid = {arXiv:2308.04851},
        pages = {arXiv:2308.04851},
          doi = {10.48550/arXiv.2308.04851},
archivePrefix = {arXiv},
       eprint = {2308.04851},
 primaryClass = {astro-ph.SR},
       adsurl = {https://ui.adsabs.harvard.edu/abs/2023arXiv230804851T}
}

@ARTICLE{trellesarjona021,
       author = {{Trelles Arjona}, J.~C. and {Mart{\'\i}nez Gonz{\'a}lez}, M.~J. and {Ruiz Cobo}, B.},
        title = "{Mapping the Hidden Magnetic Field of the Quiet Sun}",
      journal = {\apjl},
         year = 2021,
        month = jul,
       volume = {915},
       number = {1},
          eid = {L20},
        pages = {L20},
          doi = {10.3847/2041-8213/ac0af2},
archivePrefix = {arXiv},
       eprint = {2106.10546},
 primaryClass = {astro-ph.SR},
       adsurl = {https://ui.adsabs.harvard.edu/abs/2021ApJ...915L..20T}
}

@ARTICLE{kasper2021alfven,
       author = {{Kasper}, J.~C. and {Klein}, K.~G. and {Lichko}, E. and {Huang}, Jia and {Chen}, C.~H.~K. and {Badman}, S.~T. and {Bonnell}, J. and {Whittlesey}, P.~L. and {Livi}, R. and {Larson}, D. and {Pulupa}, M. and {Rahmati}, A. and {Stansby}, D. and {Korreck}, K.~E. and {Stevens}, M. and {Case}, A.~W. and {Bale}, S.~D. and {Maksimovic}, M. and {Moncuquet}, M. and {Goetz}, K. and {Halekas}, J.~S. and {Malaspina}, D. and {Raouafi}, Nour E. and {Szabo}, A. and {MacDowall}, R. and {Velli}, Marco and {Dudok de Wit}, Thierry and {Zank}, G.~P.},
        title = "{Parker Solar Probe Enters the Magnetically Dominated Solar Corona}",
      journal = {\prl},
         year = 2021,
        month = dec,
       volume = {127},
       number = {25},
          eid = {255101},
        pages = {255101},
          doi = {10.1103/PhysRevLett.127.255101},
       adsurl = {https://ui.adsabs.harvard.edu/abs/2021PhRvL.127y5101K}
}

@ARTICLE{perozzi2017,
       author = {{Perozzi}, Ettore and {Ceccaroni}, Marta and {Valsecchi}, Giovanni B. and {Rossi}, Alessandro},
        title = "{Distant retrograde orbits and the asteroid hazard}",
      journal = {European Physical Journal Plus},
         year = 2017,
        month = aug,
       volume = {132},
       number = {8},
          eid = {367},
        pages = {367},
          doi = {10.1140/epjp/i2017-11644-0},
       adsurl = {https://ui.adsabs.harvard.edu/abs/2017EPJP..132..367P}
}

@ARTICLE{davies2026_solo,
       author = {{Davies}, Emma E. and {Weiler}, Eva and {M{\"o}stl}, Christian and {Horbury}, Timothy S. and {O'Brien}, Helen and {Morris}, Jean and {Crabtree}, Alastair},
        title = "{Real-time prediction of geomagnetic storms using Solar Orbiter as a far upstream solar wind monitor}",
      journal = {arXiv e-prints},
         year = 2026,
        month = feb,
          eid = {arXiv:2508.13892},
        pages = {arXiv:2508.13892},
          doi = {10.48550/arXiv.2508.13892},
archivePrefix = {arXiv},
       eprint = {2508.13892},
 primaryClass = {physics.space-ph},
       adsurl = {https://ui.adsabs.harvard.edu/abs/2025arXiv250813892D}
}

@ARTICLE{kooi2022_faraday,
       author = {{Kooi}, Jason E. and {Wexler}, David B. and {Jensen}, Elizabeth A. and {Kenny}, Megan N. and {Nieves-Chinchilla}, Teresa and {Wilson}, III, Lynn B. and {Wood}, Brian E. and {Jian}, Lan K. and {Fung}, Shing F. and {Pevtsov}, Alexei and {Gopalswamy}, Nat and {Manchester}, Ward B.},
        title = "{Modern Faraday Rotation Studies to Probe the Solar Wind}",
      journal = {Frontiers in Astronomy and Space Sciences},
         year = 2022,
        month = apr,
       volume = {9},
          eid = {841866},
        pages = {841866},
          doi = {10.3389/fspas.2022.841866},
       adsurl = {https://ui.adsabs.harvard.edu/abs/2022FrASS...941866K}
}

@article{kay2026collection,
  author = {Kay, C. and Davies, E. E. and Dumbović, M. and Martinić, K. and Palmerio, E. and Rüdisser, H. T. and Weiler, E. and Möstl, C.},
  title = {Collection, Collation, and Comparison of Near-Earth In Situ {CME} Boundaries},
  journal = {Space Weather, in revision},
  year = {2026},
  note = {submitted}
}

@ARTICLE{jensen2025polarized,
       author = {{Jensen}, Elizabeth A. and {Manchester}, W.~B. and {Kooi}, J.~E. and {Nieves-Chinchilla}, T. and {Jian}, L.~K. and {Wexler}, D.~B. and {Fung}, S.~F. and {Gopalswamy}, N.},
        title = "{Modeling Polarized Radio Sounding Observations of a Coronal Mass Ejection}",
      journal = {\apj},
         year = 2025,
        month = jul,
       volume = {987},
       number = {2},
          eid = {156},
        pages = {156},
          doi = {10.3847/1538-4357/add1c3},
archivePrefix = {arXiv},
       eprint = {2209.03350},
 primaryClass = {astro-ph.IM},
       adsurl = {https://ui.adsabs.harvard.edu/abs/2025ApJ...987..156J}
}

@ARTICLE{wood2020faraday,
       author = {{Wood}, Brian E. and {Tun-Beltran}, Samuel and {Kooi}, Jason E. and {Polisensky}, Emil J. and {Nieves-Chinchilla}, Teresa},
        title = "{Inferences About the Magnetic Field Structure of a CME with Both In Situ and Faraday Rotation Constraints}",
      journal = {\apj},
         year = 2020,
        month = jun,
       volume = {896},
       number = {2},
          eid = {99},
        pages = {99},
          doi = {10.3847/1538-4357/ab93b8},
archivePrefix = {arXiv},
       eprint = {2006.10794},
 primaryClass = {astro-ph.SR},
       adsurl = {https://ui.adsabs.harvard.edu/abs/2020ApJ...896...99W}
}

@ARTICLE{ruedisser2026arcane,
       author = {{R{\"u}disser}, H.~T. and {Nguyen}, G. and {Le Lou{\"e}dec}, J. and {Davies}, E.~E. and {M{\"o}stl}, C.},
        title = "{ARCANE -- Early Detection of Interplanetary Coronal Mass Ejections}",
      journal = {Space Weather},
         year = 2026,
        month = feb,               
        volume = {24},
       number = {2},
        pages = {2025SW004537},
          doi = {10.1029/2025SW004537},
archivePrefix = {arXiv},
       eprint = {2505.09365},
 primaryClass = {physics.space-ph},
       adsurl = {https://ui.adsabs.harvard.edu/abs/2025arXiv250509365R}
}

@ARTICLE{ruohotie2025,
       author = {{Ruohotie}, Julia and {Good}, Simon and {M{\"o}stl}, Christian and {Kilpua}, Emilia},
        title = "{Intermittency in Interplanetary Coronal Mass Ejections Observed by Parker Solar Probe and Solar Orbiter}",
      journal = {\apjl},
         year = 2025,
        month = jun,
       volume = {986},
       number = {2},
          eid = {L27},
        pages = {L27},
          doi = {10.3847/2041-8213/ade0b0},
archivePrefix = {arXiv},
       eprint = {2505.22283},
 primaryClass = {astro-ph.SR},
       adsurl = {https://ui.adsabs.harvard.edu/abs/2025ApJ...986L..27R}
}

@ARTICLE{henon1969,
       author = {{Henon}, M.},
        title = "{Numerical exploration of the restricted problem, V}",
      journal = {\aap},
         year = 1969,
        month = feb,
       volume = {1},
        pages = {223-238},
       adsurl = {https://ui.adsabs.harvard.edu/abs/1969A&A.....1..223H}
}

@ARTICLE{weiss2024distorted,
       author = {{Weiss}, Andreas J. and {Nieves-Chinchilla}, Teresa and {M{\"o}stl}, Christian},
        title = "{Distorted Magnetic Flux Ropes within Interplanetary Coronal Mass Ejections}",
      journal = {\apj},
         year = 2024,
        month = nov,
       volume = {975},
       number = {2},
          eid = {169},
        pages = {169},
          doi = {10.3847/1538-4357/ad7940},
archivePrefix = {arXiv},
       eprint = {2406.13022},
 primaryClass = {astro-ph.SR},
       adsurl = {https://ui.adsabs.harvard.edu/abs/2024ApJ...975..169W}
}

@ARTICLE{trotta2024,
       author = {{Trotta}, Domenico and {Larosa}, Andrea and {Nicolaou}, Georgios and {Horbury}, Timothy S. and {Matteini}, Lorenzo and {Hietala}, Heli and {Blanco-Cano}, Xochitl and {Franci}, Luca and {Chen}, C.~H.~K. and {Zhao}, Lingling and {Zank}, Gary P. and {Cohen}, Christina M.~S. and {Bale}, Stuart D. and {Laker}, Ronan and {Fargette}, Nais and {Valentini}, Francesco and {Khotyaintsev}, Yuri and {Kieokaew}, Rungployphan and {Raouafi}, Nour and {Davies}, Emma and {Vainio}, Rami and {Dresing}, Nina and {Kilpua}, Emilia and {Karlsson}, Tomas and {Owen}, Christopher J. and {Wimmer-Schweingruber}, Robert F.},
        title = "{Properties of an Interplanetary Shock Observed at 0.07 and 0.7 au by Parker Solar Probe and Solar Orbiter}",
      journal = {\apj},
         year = 2024,
        month = feb,
       volume = {962},
       number = {2},
          eid = {147},
        pages = {147},
          doi = {10.3847/1538-4357/ad187d},
archivePrefix = {arXiv},
       eprint = {2312.05983},
 primaryClass = {astro-ph.SR},
       adsurl = {https://ui.adsabs.harvard.edu/abs/2024ApJ...962..147T}
}

@ARTICLE{cyr2000diamond,
       author = {{St. Cyr}, O.~C. and {Mesarch}, M.~A. and {Maldonado}, H.~M. and {Folta}, D.~C. and {Harper}, A.~D. and {Davila}, J.~M. and {Fisher}, R.~R.},
        title = "{Space Weather Diamond: a four spacecraft monitoring system}",
      journal = {Journal of Atmospheric and Solar-Terrestrial Physics},
         year = 2000,
        month = sep,
       volume = {62},
       number = {14},
        pages = {1251-1255},
          doi = {10.1016/S1364-6826(00)00069-9},
       adsurl = {https://ui.adsabs.harvard.edu/abs/2000JASTP..62.1251S}
}

@ARTICLE{cicalo2025henon,
       author = {{Cical{\`o}}, Stefano and {Alessi}, Elisa Maria and {Provinciali}, Lorenzo and {Amabili}, Paride and {Saita}, Giorgio and {Calcagno}, Davide and {Marcucci}, Maria Federica and {Laurenza}, Monica and {Zimbardo}, Gaetano and {Landi}, Simone and {Walker}, Roger and {Khan}, Michael},
        title = "{Mission analysis for the HENON CubeSat mission to a large Sun-Earth distant retrograde orbit}",
      journal = {\apss},
         year = 2025,
        month = aug,
       volume = {370},
       number = {8},
          eid = {83},
        pages = {83},
          doi = {10.1007/s10509-025-04473-0},
archivePrefix = {arXiv},
       eprint = {2508.02138},
 primaryClass = {astro-ph.EP},
       adsurl = {https://ui.adsabs.harvard.edu/abs/2025Ap&SS.370...83C}
}

@ARTICLE{weiler2025,
       author = {{Weiler}, E. and {M{\"o}stl}, C. and {Davies}, E.~E. and {Veronig}, A.~M. and {Amerstorfer}, U.~V. and {Amerstorfer}, T. and {Le Lou{\"e}dec}, J. and {Bauer}, M. and {Lugaz}, N. and {Haberle}, V. and {R{\"u}disser}, H.~T. and {Majumdar}, S. and {Reiss}, M.},
        title = "{First Observations of a Geomagnetic Superstorm With a Sub-L1 Monitor}",
      journal = {Space Weather},
         year = 2025,
        month = mar,
       volume = {23},
       number = {3},
        pages = {2024SW004260},
          doi = {10.1029/2024SW004260},
archivePrefix = {arXiv},
       eprint = {2411.12490},
 primaryClass = {physics.space-ph},
       adsurl = {https://ui.adsabs.harvard.edu/abs/2025SpWea..2304260W}
}

@ARTICLE{lugaz2024ssr,
       author = {{Lugaz}, No{\'e} and {Lee}, Christina O. and {Al-Haddad}, Nada and {Lillis}, Robert J. and {Jian}, Lan K. and {Curtis}, David W. and {Galvin}, Antoinette B. and {Whittlesey}, Phyllis L. and {Rahmati}, Ali and {Zesta}, Eftyhia and {Moldwin}, Mark and {Summerlin}, Errol J. and {Larson}, Davin E. and {Courtade}, Sasha and {French}, Richard and {Hunter}, Richard and {Covitti}, Federico and {Cosgrove}, Daniel and {Prall}, J.~D. and {Allen}, Robert C. and {Zhuang}, Bin and {Winslow}, R{\'e}ka M. and {Scolini}, Camilla and {Lynch}, Benjamin J. and {Filwett}, Rachael J. and {Palmerio}, Erika and {Farrugia}, Charles J. and {Smith}, Charles W. and {M{\"o}stl}, Christian and {Weiler}, Eva and {Janvier}, Miho and {Regnault}, Florian and {Livi}, Roberto and {Nieves-Chinchilla}, Teresa},
        title = "{The Need for Near-Earth Multi-Spacecraft Heliospheric Measurements and an Explorer Mission to Investigate Interplanetary Structures and Transients in the Near-Earth Heliosphere}",
      journal = {\ssr},
         year = 2024,
        month = oct,
       volume = {220},
       number = {7},
          eid = {73},
        pages = {73},
          doi = {10.1007/s11214-024-01108-8},
       adsurl = {https://ui.adsabs.harvard.edu/abs/2024SSRv..220...73L}
}

@ARTICLE{lugaz2025need,
       author = {{Lugaz}, No{\'e} and {Al-Haddad}, Nada and {Zhuang}, Bin and {M{\"o}stl}, Christian and {Davies}, Emma E. and {Farrugia}, Charles J. and {Banu}, Sahanaj Aktar and {Weiler}, Eva and {Galvin}, Antoinette B.},
        title = "{The Need for a Sub-L1 Space Weather Research Mission: Current Knowledge Gaps on Coronal Mass Ejections}",
      journal = {Space Weather},
         year = 2025,
        month = feb,
       volume = {23},
       number = {2},
        pages = {2024SW004189},
          doi = {10.1029/2024SW004189},
       adsurl = {https://ui.adsabs.harvard.edu/abs/2025SpWea..2304189L}
}

@ARTICLE{alhaddad2025structure,
       author = {{Al-Haddad}, Nada and {Lugaz}, No{\'e}},
        title = "{The Magnetic Field Structure of Coronal Mass Ejections: A More Realistic Representation}",
      journal = {\ssr},
         year = 2025,
        month = feb,
       volume = {221},
       number = {1},
          eid = {12},
        pages = {12},
          doi = {10.1007/s11214-025-01138-w},
       adsurl = {https://ui.adsabs.harvard.edu/abs/2025SSRv..221...12A}
}

@ARTICLE{palmerio2025event4,
       author = {{Palmerio}, Erika and {Kay}, Christina and {Al-Haddad}, Nada and {Lynch}, Benjamin J. and {Trotta}, Domenico and {Yu}, Wenyuan and {Ledvina}, Vincent E. and {S{\'a}nchez-Cano}, Beatriz and {Riley}, Pete and {Heyner}, Daniel and {Schmid}, Daniel and {Fischer}, David and {Richter}, Ingo and {Auster}, Hans-Ulrich},
        title = "{A coronal mass ejection encountered by four spacecraft within 1 au from the Sun: ensemble modelling of propagation and magnetic structure}",
      journal = {\mnras},
         year = 2025,
        month = jan,
       volume = {536},
       number = {1},
        pages = {203-222},
          doi = {10.1093/mnras/stae2606},
archivePrefix = {arXiv},
       eprint = {2411.12706},
 primaryClass = {astro-ph.SR},
       adsurl = {https://ui.adsabs.harvard.edu/abs/2025MNRAS.536..203P}
}

@ARTICLE{regnault2024,
       author = {{Regnault}, F. and {Al-Haddad}, N. and {Lugaz}, N. and {Farrugia}, C.~J. and {Yu}, W. and {Zhuang}, B. and {Davies}, E.~E.},
        title = "{Discrepancies in the Properties of a Coronal Mass Ejection on Scales of 0.03 au as Revealed by Simultaneous Measurements at Solar Orbiter and Wind: The 2021 November 3{\textendash}5 Event}",
      journal = {\apj},
         year = 2024,
        month = feb,
       volume = {962},
       number = {2},
          eid = {190},
        pages = {190},
          doi = {10.3847/1538-4357/ad1883},
archivePrefix = {arXiv},
       eprint = {2311.14046},
 primaryClass = {astro-ph.SR},
       adsurl = {https://ui.adsabs.harvard.edu/abs/2024ApJ...962..190R}
}

@ARTICLE{ruedisser_2024,
       author = {{R{\"u}disser}, Hannah T. and {Weiss}, Andreas J. and {Le Lou{\"e}dec}, Justin and {Amerstorfer}, Ute V. and {M{\"o}stl}, Christian and {Davies}, Emma E. and {Lammer}, Helmut},
        title = "{Understanding the Effects of Spacecraft Trajectories through Solar Coronal Mass Ejection Flux Ropes Using 3DCOREweb}",
      journal = {\apj},
         year = 2024,
        month = oct,
       volume = {973},
       number = {2},
          eid = {150},
        pages = {150},
          doi = {10.3847/1538-4357/ad660a},
archivePrefix = {arXiv},
       eprint = {2405.03271},
 primaryClass = {astro-ph.SR},
       adsurl = {https://ui.adsabs.harvard.edu/abs/2024ApJ...973..150R}
}

@ARTICLE{pal2022_model,
       author = {{Pal}, Sanchita and {Nandy}, Dibyendu and {Kilpua}, Emilia K.~J.},
        title = "{Magnetic cloud prediction model for forecasting space weather relevant properties of Earth-directed coronal mass ejections}",
      journal = {\aap},
         year = 2022,
        month = sep,
       volume = {665},
          eid = {A110},
        pages = {A110},
          doi = {10.1051/0004-6361/202243513},
archivePrefix = {arXiv},
       eprint = {2203.05231},
 primaryClass = {astro-ph.SR},
       adsurl = {https://ui.adsabs.harvard.edu/abs/2022A&A...665A.110P}
}

@ARTICLE{davies2021_juno,
       author = {{Davies}, Emma E. and {Forsyth}, Robert J. and {Winslow}, R{\'e}ka M. and {M{\"o}stl}, Christian and {Lugaz}, No{\'e}},
        title = "{A Catalog of Interplanetary Coronal Mass Ejections Observed by Juno between 1 and 5.4 au}",
      journal = {\apj},
         year = 2021,
        month = dec,
       volume = {923},
       number = {2},
          eid = {136},
        pages = {136},
          doi = {10.3847/1538-4357/ac2ccb},
archivePrefix = {arXiv},
       eprint = {2111.11336},
 primaryClass = {physics.space-ph},
       adsurl = {https://ui.adsabs.harvard.edu/abs/2021ApJ...923..136D}
}

@ARTICLE{richardson2014identification,
       author = {{Richardson}, I.~G.},
        title = "{Identification of Interplanetary Coronal Mass Ejections at Ulysses Using Multiple Solar Wind Signatures}",
      journal = {\solphys},
         year = 2014,
        month = oct,
       volume = {289},
       number = {10},
        pages = {3843-3894},
          doi = {10.1007/s11207-014-0540-8},
       adsurl = {https://ui.adsabs.harvard.edu/abs/2014SoPh..289.3843R}
}

@ARTICLE{laker2024,
       author = {{Laker}, R. and {Horbury}, T.~S. and {O'Brien}, H. and {Fauchon-Jones}, E.~J. and {Angelini}, V. and {Fargette}, N. and {Amerstorfer}, T. and {Bauer}, M. and {M{\"o}stl}, C. and {Davies}, E.~E. and {Davies}, J.~A. and {Harrison}, R. and {Barnes}, D. and {Dumbovi{\'c}}, M.},
        title = "{Using Solar Orbiter as an Upstream Solar Wind Monitor for Real Time Space Weather Predictions}",
      journal = {Space Weather},
         year = 2024,
        month = feb,
       volume = {22},
       number = {2},
          eid = {e2023SW003628},
        pages = {e2023SW003628},
          doi = {10.1029/2023SW003628},
archivePrefix = {arXiv},
       eprint = {2307.01083},
 primaryClass = {physics.space-ph},
       adsurl = {https://ui.adsabs.harvard.edu/abs/2024SpWea..2203628L}
}

@ARTICLE{camporeale2017,
       author = {{Camporeale}, Enrico and {Car{\`e}}, Algo and {Borovsky}, Joseph E.},
        title = "{Classification of Solar Wind With Machine Learning}",
      journal = {Journal of Geophysical Research (Space Physics)},
         year = 2017,
        month = nov,
       volume = {122},
       number = {11},
        pages = {10,910-10,920},
          doi = {10.1002/2017JA024383},
archivePrefix = {arXiv},
       eprint = {1710.02313},
 primaryClass = {physics.space-ph},
       adsurl = {https://ui.adsabs.harvard.edu/abs/2017JGRA..12210910C}
}

@ARTICLE{nguyen2019,
       author = {{Nguyen}, Gautier and {Aunai}, Nicolas and {Fontaine}, Dominique and {Le Pennec}, Erwan and {Van den Bossche}, Joris and {Jeandet}, Alexis and {Bakkali}, Brice and {Vignoli}, Louis and {Regaldo-Saint Blancard}, Bruno},
        title = "{Automatic Detection of Interplanetary Coronal Mass Ejections from In Situ Data: A Deep Learning Approach}",
      journal = {\apj},
         year = 2019,
        month = apr,
       volume = {874},
       number = {2},
          eid = {145},
        pages = {145},
          doi = {10.3847/1538-4357/ab0d24},
archivePrefix = {arXiv},
       eprint = {1903.10780},
 primaryClass = {astro-ph.SR},
       adsurl = {https://ui.adsabs.harvard.edu/abs/2019ApJ...874..145N}
}

@ARTICLE{nguyen2025_multiclass,
       author = {{Nguyen}, Gautier and {Bernoux}, Guillerme and {Ferlin}, Antoine},
        title = "{Simultaneous multi-class detection of interplanetary space weather events}",
      journal = {Journal of Space Weather and Space Climate},
         year = 2025,
        month = jan,
       volume = {15},
          eid = {21},
        pages = {21},
          doi = {10.1051/swsc/2025016},
       adsurl = {https://ui.adsabs.harvard.edu/abs/2025JSWSC..15...21N}
}

@ARTICLE{ruedisser2022,
       author = {{R{\"u}disser}, H.~T. and {Windisch}, A. and {Amerstorfer}, U.~V. and {M{\"o}stl}, C. and {Amerstorfer}, T. and {Bailey}, R.~L. and {Reiss}, M.~A.},
        title = "{Automatic Detection of Interplanetary Coronal Mass Ejections in Solar Wind In Situ Data}",
      journal = {Space Weather},
         year = 2022,
        month = oct,
       volume = {20},
       number = {10},
          eid = {e2022SW003149},
        pages = {e2022SW003149},
          doi = {10.1029/2022SW003149},
archivePrefix = {arXiv},
       eprint = {2205.03578},
 primaryClass = {astro-ph.SR},
       adsurl = {https://ui.adsabs.harvard.edu/abs/2022SpWea..2003149R}
}

@ARTICLE{moestl2022,
       author = {{M{\"o}stl}, Christian and {Weiss}, Andreas J. and {Reiss}, Martin A. and {Amerstorfer}, Tanja and {Bailey}, Rachel L. and {Hinterreiter}, J{\"u}rgen and {Bauer}, Maike and {Barnes}, David and {Davies}, Jackie A. and {Harrison}, Richard A. and {Freiherr von Forstner}, Johan L. and {Davies}, Emma E. and {Heyner}, Daniel and {Horbury}, Tim and {Bale}, Stuart D.},
        title = "{Multipoint Interplanetary Coronal Mass Ejections Observed with Solar Orbiter, BepiColombo, Parker Solar Probe, Wind, and STEREO-A}",
      journal = {\apjl},
         year = 2022,
        month = jan,
       volume = {924},
       number = {1},
          eid = {L6},
        pages = {L6},
          doi = {10.3847/2041-8213/ac42d0},
archivePrefix = {arXiv},
       eprint = {2109.07200},
 primaryClass = {astro-ph.SR},
       adsurl = {https://ui.adsabs.harvard.edu/abs/2022ApJ...924L...6M}
}

@ARTICLE{sarkar2024,
       author = {{Sarkar}, Ranadeep and {Srivastava}, Nandita and {Gopalswamy}, Nat and {Kilpua}, Emilia},
        title = "{Modeling the Magnetic Vectors of Interplanetary Coronal Mass Ejections at Different Heliocentric Distances with INFROS}",
      journal = {\apjs},
         year = 2024,
        month = aug,
       volume = {273},
       number = {2},
          eid = {36},
        pages = {36},
          doi = {10.3847/1538-4365/ad5835},
archivePrefix = {arXiv},
       eprint = {2406.09247},
 primaryClass = {astro-ph.SR},
       adsurl = {https://ui.adsabs.harvard.edu/abs/2024ApJS..273...36S}
}

@ARTICLE{davies2022,
       author = {{Davies}, Emma E. and {Winslow}, R{\'e}ka M. and {Scolini}, Camilla and {Forsyth}, Robert J. and {M{\"o}stl}, Christian and {Lugaz}, No{\'e} and {Galvin}, Antoinette B.},
        title = "{Multi-spacecraft Observations of the Evolution of Interplanetary Coronal Mass Ejections between 0.3 and 2.2 au: Conjunctions with the Juno Spacecraft}",
      journal = {\apj},
         year = 2022,
        month = jul,
       volume = {933},
       number = {2},
          eid = {127},
        pages = {127},
          doi = {10.3847/1538-4357/ac731a},
archivePrefix = {arXiv},
       eprint = {2205.09472},
 primaryClass = {physics.space-ph},
       adsurl = {https://ui.adsabs.harvard.edu/abs/2022ApJ...933..127D}
}
\bibliographystyle{aasjournalv7}


\end{document}